\documentclass[12pt,letterpaper,titlepage,reqno,oneside]{article}
\usepackage[utf8]{inputenc}
\usepackage{graphicx}
\usepackage[super,comma]{natbib}
\usepackage[english]{babel}
\usepackage[figurename=FIG.,tablename=TABLE]{caption}
\usepackage{enumerate}
\usepackage{amsmath,amsfonts,amssymb,wasysym,latexsym,amsthm,bm}
\usepackage[mathscr]{eucal}
\usepackage{color}
\usepackage{subfig}
\usepackage{siunitx}
\usepackage{textcomp}
\usepackage{titlesec}
\usepackage[subfigure,titles]{tocloft}
\usepackage{setspace}

\DeclareCaptionFormat{upper}{#1#2\uppercase{#3}\par}

\titleformat{\section}{\normalsize\bfseries}{\Roman{section}.}{1em}{\normalsize\uppercase}
\titleformat{\subsection}{\normalsize\bfseries}{\Alph{subsection}.}{1em}{}

\renewcommand{\thesection}{\Roman{section}}

\cftpagenumbersoff{figure}
\cftpagenumbersoff{subfigure}

\makeatletter
\renewcommand\listoffigures{%
    \medskip\raggedright \textbf{LIST OF FIGURES}%
    \@mkboth{\MakeUppercase\listfigurename}%
        {\MakeUppercase\listfigurename}%
    \@starttoc{lof}%
}
\makeatother

\makeatletter
\def\p@subsection{\thesection.\,}
\makeatother

\begin{document}
\sisetup{detect-family,detect-display-math=true,exponent-product=\cdot,output-complex-root=\text{\ensuremath{i}},per-mode=symbol}

\author{%
Joris P. Oosterhuis\footnote{Author to whom correspondence should be addressed. E-mail: j.p.oosterhuis@utwente.nl}, Simon B\"uhler, and Theo H. van der Meer \\
\textit{Department of Thermal Engineering, University of Twente,}\\
\textit{Enschede, The Netherlands}\\ \\
Douglas Wilcox\\
\textit{Chart Inc., Troy, New York}%
}
\title{\LARGE{A numerical investigation on the vortex formation and flow separation of the oscillatory flow in jet pumps} \\
\textit{\large{Oscillatory flow fields in jet pumps}}
}
\date{\today}

\maketitle

\begin{abstract}
A two-dimensional computational fluid dynamics model is used to predict the oscillatory flow through a tapered cylindrical tube section (jet pump) placed in a larger outer tube. Due to the shape of the jet pump, there will exist an asymmetry in the hydrodynamic end effects which will cause a time-averaged pressure drop to occur that can be used to cancel Gedeon streaming in a closed-loop thermoacoustic device. The performance of two jet pump geometries with different taper angles is investigated. A specific time-domain impedance boundary condition is implemented in order to simulate traveling acoustic wave conditions. It is shown that by scaling the acoustic displacement amplitude to the jet pump dimensions, similar minor losses are observed independent of the jet pump geometry. Four different flow regimes are distinguished and the observed flow phenomena are related to the jet pump performance. The simulated jet pump performance is compared to an existing quasi-steady approximation which is shown to only be valid for small displacement amplitudes compared to the jet pump length.\\



\end{abstract}

\addtocounter{page}{2}
\section{Introduction}

\begin{figure}
\centering
\includegraphics[width=.5\textwidth]{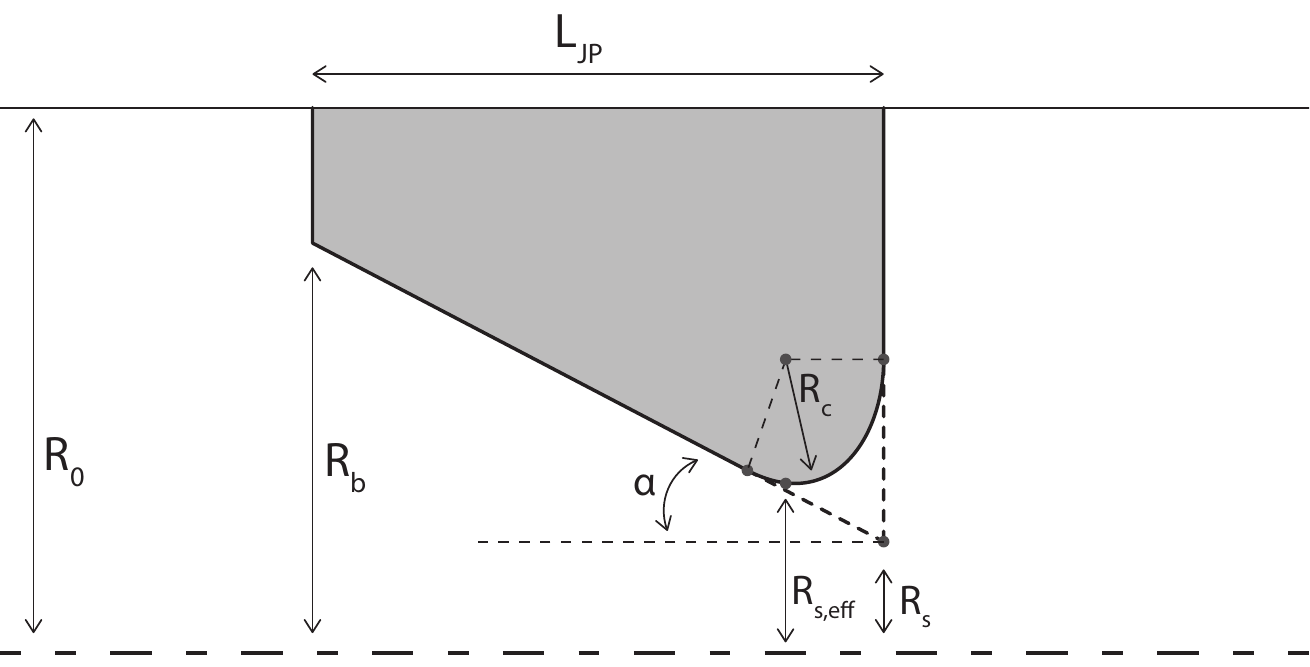}
\caption{Jet pump with parameters that define the geometry (not to scale). Bottom dashed line indicates centerline, top solid line indicates tube wall.}
\label{fig:jetpumpgeom}
\end{figure}

A jet pump is a crucial part of most closed-loop thermoacoustic devices.\cite{Backhaus1999} In such devices, a time-averaged mass flux known as Gedeon streaming can exist.\cite{Gedeon1997} This time-averaged mass flux results in convective heat transport that can severely degrade the efficiency of thermoacoustic devices.\cite{Swift1999} To suppress Gedeon streaming, a jet pump can be used. Backhaus and Swift have shown that by correctly shaping a jet pump it is possible to take advantage of asymmetric hydrodynamic end effects to impose a pressure drop across the jet pump.\cite{Backhaus2000}  A typical jet pump consists of a narrowed, tapered tube section as shown schematically in Fig.~\ref{fig:jetpumpgeom}. By balancing the pressure drop across the jet pump with that which exists across the regenerator, it is possible to produce a net zero time-averaged mass flux in the thermoacoustic device.

Despite the proven effectiveness of jet pumps, there is a lack of understanding with respect to the exact fluid dynamics that lead to the observed pressure drop. Current criteria for the design of a jet pump assume that the flow at any point in time has little ``memory'' of its past history | which is often referred to as the Iguchi-hypothesis.\cite{Iguchi1982_u-shaped} This allows the acoustic behavior to be based on a quasi-steady approximation using minor loss coefficients reported for steady pipe flow.\cite{Swift1999}

\subsection{Quasi-steady theory}

The pressure drop generated by an abrupt pipe transition in steady flow can be calculated using
\begin{equation}
\Delta p_{ml} =  \frac{1}{2} K\rho u^2,
\label{eq:minorloss}
\end{equation}
where $K$ is the minor loss coefficient which depends on geometry and flow direction, $\rho$ is the fluid density and $u$ is the fluid velocity. For an abrupt expansion, $K=K_\mathit{exp}$ and can be estimated using the Borda-Carnot equation.\cite{Idelchik2007} In the case of a uniform flow
\begin{equation}
K_\mathit{exp} = \left(1-\frac{A_s}{A_0}\right)^2,
\label{eq:bordacarnot}
\end{equation}
where $A_s$ is the smaller cross-sectional area before the expansion and $A_0$ is the cross-sectional area right after the expansion. Note that these values assume a uniform velocity profile. Non-uniform velocity profiles will result in larger minor loss coefficient values for expansion.\cite{Wakeland2002} 

For a contraction, the steady flow minor loss coefficient is dependent upon the dimensionless curvature of the transition, $R_c/D$, where $R_c$ is the radius of curvature and $D$ is the diameter of the opening.\cite{Idelchik2007} For a sharp contraction, $K_\mathit{con}=0.5$ but this reduces to $K_\mathit{con}=0.04$ for $R_c/D \geq 0.15$.

Under the assumption that the Iguchi-hypothesis is applicable and that the minor loss coefficients have the same values in oscillatory flow as they do in steady flow, a quasi-steady model has been formulated by Backhaus and Swift to calculate the time-averaged pressure drop across a jet pump\cite{Backhaus2000}
\begin{equation}
\Delta p_{2,\mathit{JP}}  = \frac{1}{8}\rho_0 |u_{1,\mathit{JP}}|^2\left[(K_{\mathit{exp},s}-K_{\mathit{con},s}) + \left(\frac{A_s}{A_b}\right)^2(K_{\mathit{con},b}-K_{\mathit{exp},b})\right],
\label{eq:backhaus}
\end{equation}
where $|u_{1,\mathit{JP}}|$ is the velocity amplitude at the small exit of the jet pump. The subscripts ``$s$'' and ``$b$'' indicate the small and big opening of the jet pump, respectively.

Although this time-averaged pressure drop can be exploited to cancel Gedeon streaming and improve the efficiency of a looped thermoacoustic device, this approach is not without penalty. Adding a jet pump results in additional dissipation of acoustic power. Under the same previous assumptions, the time-averaged acoustic power dissipation across a jet pump is~\cite{Backhaus2000}
\begin{equation}
\Delta \dot{E}_\mathit{JP} = \frac{\rho_0 |u_{1,\mathit{JP}}|^3A_s}{3\pi}\left[(K_{\mathit{exp},s}+K_{\mathit{con},s}) + \left(\frac{A_s}{A_b}\right)^2(K_{\mathit{con},b}+K_{\mathit{exp},b})\right].
\label{eq:backhaus_dE}
\end{equation}

Qualitative evidence exists which supports the current analysis, but quantitative agreement between the theory and experiments remains poor.\cite{Backhaus2000,Petculescu2003} While the accuracy of this approach is yet unknown, it is assumed valid for large displacement amplitudes in relation to the jet pump dimensions.\cite{Backhaus2000}
Moreover, when using minor loss coefficients for steady expansion and contraction, the effect of the jet pump taper angle or the jet pump length is not included in the current theory while it is observed to have an important effect on the jet pump pressure drop.\cite{Petculescu2003,Oosterhuis2014}

\subsection{Literature review}

Previous studies related to jet pumps for thermoacoustic applications include mainly experimental or applied work; only a few computational studies have been published to date. Petculescu and Wilen measured the pressure drop for a series of jet pump geometries in a standing wave experimental apparatus.\cite{Petculescu2003} They then derived minor loss coefficients based on the measured pressure and the velocity in the jet pump waist which was estimated using an acoustic network model. A difference between the measured and theoretical minor loss coefficients is reported, especially for the diverging flow direction. Nevertheless, for the investigated geometries -- up to a taper angle of \SI{10}{\degree} -- good agreement between the performed steady flow and oscillating experiments is obtained. An increase in the taper angle is shown to have a negative influence on the time-averaged pressure drop.

Smith and Swift have experimentally studied oscillatory flow through a nozzle with constant diameter, simulating one end of a jet pump.\cite{Smith2003a} In their work, a nozzle is connected to open space, establishing a non-confined jet. A parametric study on the time-averaged pressure drop and the acoustic power dissipation is performed, identifying some of the dimensionless quantities which describe the flow phenomena: the dimensionless stroke length, the dimensionless curvature and the acoustic Reynolds number. Furthermore, a Schlieren visualization of the flow field is presented. The formation of a vortex pair and a turbulent jet is observed. It is concluded that ``extensive numerical studies'' are required for a further understanding of the minor loss phenomena to control streaming. In a separate article,\cite{Smith2003b} Smith and Swift compare the characteristics of a synthetic (oscillatory) jet to a continuous jet in the same experimental setup. The self-similar velocity profiles are found to be identical but the jet width of the synthetic jets grow more rapidly than the continuous jets.

Computational studies related to jet pumps mainly include the work of Boluriaan and Morris.\cite{Morris2001,Morris2004} In two studies, the minor losses due to a single diameter transition under standing wave conditions are simulated using a two-dimensional computational fluid dynamics (CFD) model. The standing wave is generated by either applying an oscillatory body force (``shaking'' the domain) or by using an oscillatory line source inside the domain. Axial pressure and velocity profiles are presented and the effect of jetting and vortex shedding on the flow field is described. The time-averaged pressure drop across the transition is found to be a factor of three higher than the quasi-steady solution. In a separate study, a jet pump geometry is investigated using a similar CFD model.\cite{Boluriaan2003a} In this case, a combination of two line sources with a non-reflecting boundary condition on either side is used to generate a traveling wave inside the domain. The flow field is calculated for a single jet pump geometry and wave amplitude.

The authors contribution to the field is limited to a preliminary study~\cite{Oosterhuis2014} where the effect of the jet pump taper angle on the time-averaged pressure drop is investigated numerically and compared against the experimental work of Petculescu and Wilen.\cite{Petculescu2003} A clear decrease in time-averaged pressure drop is observed at higher taper angles, which is one of the motivations for the work presented here.\\

In this paper, the oscillatory flow in the vicinity of a jet pump is investigated using a CFD model which is described in Section~\ref{sec:modeling}. Using this CFD model, the performance of two jet pump geometries with different taper angles are studied. Four different flow regimes are described (Section~\ref{sec:flow_regimes}) and subsequently linked to the observed jet pump performance. The time-averaged pressure drop and acoustic power dissipation are scaled to relate the behavior of the two different taper angles and a comparison with the quasi-steady approximation is made (Section~\ref{sec:jet_performance}).

\section{Modeling}
\label{sec:modeling}

A two-dimensional axisymmetric CFD model is developed using the commercial software package ANSYS CFX version 14.5,\cite{ANSYS2011} which has been used successfully in the simulation of various (thermo)acoustic applications.\cite{Aben2010,Lycklama2005,Nowak2014} The jet pump is placed in an outer tube to study the influence of the jet pump geometry on the flow field. Boundary conditions are applied to simulate a traveling wave inside the computational domain; these are discussed in Sections~\ref{sec:num_setup} and~\ref{sec:TDIBC}. In all cases, air at a mean temperature of $T_0=\SI{300}{\kelvin}$ and a mean pressure of $p_0=\SI{1}{atm}$ is used as the working fluid. Three different driving frequencies are investigated: \SIlist{50;100;200}{\Hz}.

\subsection{Geometry\label{sec:geometry}}

The jet pump geometry is shown in Fig.~\ref{fig:jetpumpgeom} and is defined using a reduced number of parameters: the radius of the big exit $R_b$, the effective radius of the small exit (the jet pump ``waist'') $R_\mathit{s,eff}$, the taper half-angle $\alpha$ and the radius of curvature at the small exit of the jet pump $R_\mathit{c}$. Based on these parameters, the other parameters can be calculated. The total jet pump length $L_\mathit{JP}$ is
\begin{equation}
L_\mathit{JP} = \frac{R_b-R_s}{\tan{\alpha}},
\label{eq:Ljp}
\end{equation}
where $R_s$ is the small radius of the jet pump without any curvature applied,
\begin{equation}
R_s=R_\mathit{s,eff}-R_\mathit{c}\left(\frac{\sin\alpha+1}{\cos\alpha}-1\right).
\label{eq:Rs}
\end{equation}

In addition to the jet pump region, the computational domain consists of a section of the outer tube on both sides of the jet pump with a radius of $R_0=\SI{30}{\mm}$ and a length of $L_0=\SI{500}{\mm}$ each for the cases where $f=\SI{100}{\Hz}$. The influence of the length of this section on the jet pump performance and vortex propagation characteristics has been verified by comparing with simulations using $L_0=\SI{100}{\mm}$ and no significant difference was observed. Although a shorter outer tube section will lead to a reduced computational time, the longer length is used in order to study the resulting flow field on both sides of the jet pump in detail. For the other two driving frequencies (\SIlist{50;200}{\Hz}), $L_0$ is scaled relative to the acoustic wavelength.

Two different taper angles, \SIlist{7;15}{\degree}, are analyzed by changing the jet pump length. The corresponding jet pump lengths are shown in Table~\ref{tab:geom_varalpha}. All the other geometrical parameters remain the same and are listed in Table~\ref{tab:geom}. The dimensionless curvature is $R_c/(2R_\mathit{s,eff}) = 0.36$ for both geometries, which is well above the limit for a ``smooth'' contraction. Hence, according to steady flow literature,\cite{Idelchik2007} $K_\mathit{con}=0.04$. The ratio between the small and big cross-sectional area is $R_\mathit{s,eff}^2/R_b^2 = 0.22$ for both geometries. Because the cross-sectional area of both jet pump openings is kept constant, one would expect an identical pressure drop and acoustic power dissipation based on the quasi-steady approximation (Eq.~\ref{eq:backhaus}). Moreover, the term $(A_s/A_b)^2$ is small such that the minor losses due to the small opening of the jet pump are expected to predominantly determine the time-averaged pressure drop and acoustic power dissipation.
\begin{table}
\centering
	\caption{Jet pump length $L_\mathit{JP}$ for applied taper angles $\alpha$.}
	\label{tab:geom_varalpha}
	\begin{tabular}{ll}
	$\alpha$ & $L_\mathit{JP}$ \\
	\hline
	\SI{7}{\degree} & \SI{70.5}{\mm} \\
	\SI{15}{\degree} & \SI{35.5}{\mm} \\
	\end{tabular}
\end{table}
\begin{table}
\centering
	\caption{Dimensions of simulated jet pump geometries.}
	\label{tab:geom}
	\begin{tabular}{ll}
	$R_0$ & \SI{30}{\mm}\\
	$R_b$ & \SI{15}{\mm}\\
	$R_\mathit{s,eff}$ & \SI{7}{\mm}\\
	$R_c$ & \SI{5}{\mm}\\
	\end{tabular}
\end{table}

\subsection{Numerical setup\label{sec:num_setup}}

Within the described computational domain, the unsteady, fully compressible Navier-Stokes equations are solved. The ideal gas law is used as an equation of state whereas the energy transport is described using the total energy equation including viscous work terms.\cite{ANSYS2012_NavierStokes} No additional turbulence modeling is applied as all presented results fall within the laminar regime (see Section~\ref{sec:jet_performance}). The governing equations are discretized in space using a high resolution advection scheme and discretized in time using a second order backward Euler scheme. Each wave period is discretized using 1000 time-steps which yields a time-step size of $\Delta t=\SI{1e-5}{\s}$ for $f=\SI{100}{\Hz}$. For each simulation case, a total of $N_p=10$ wave periods are simulated. With a typical computational mesh, the total single core computational time is about 40 hours on an Intel~Core~i7 CPU.

In order to perform a two-dimensional axisymmetric simulation in ANSYS CFX, a computational mesh which extends one element in the azimuthal direction is required and symmetry boundary conditions are applied on the originating faces normal to the azimuthal direction. On the radial boundary of the outer tube (at $r=R_0$), a slip adiabatic wall boundary condition is used while at the walls of the jet pump a no-slip adiabatic wall boundary condition is imposed. 

To generate an acoustic wave, a velocity boundary condition is used at $x=0$ which oscillates in time according to $u(t) = u_1 \sin{\left(2\pi f t\right)}$ with $u_1$ a defined velocity amplitude. On the right boundary of the computational domain, at $x=L$, a dedicated time-domain impedance boundary condition~\cite{VanderPoel2013} is applied with a specified reflection coefficient of $|R|=0$, as described in Section~\ref{sec:TDIBC}. This ensures free propagation of the acoustic wave without any additional reflections being introduced into the computational domain.

\subsection{Time-domain impedance boundary condition\label{sec:TDIBC}}

A specific time-domain impedance boundary condition has been implemented in ANSYS CFX based on the work of Polifke et al.\cite{Huber2008,Kaess2008} This approach ultimately defines the pressure $p(t)$ on the boundary. The applied pressure is based on wave information from inside the domain at a previous time-step which is sampled at a distance $\Delta x = \SI{50}{\mm}$ from the boundary. Moreover, an external perturbation can be introduced on the boundary such that any complex reflection coefficient can be specified. However, in this paper only a non-reflective situation ($|R|=0$) is considered. The measured reflection coefficient typically ranges from \SIrange{1}{2}{\percent}. Further details about the exact implementation and validation can be found in the work of Van~der~Poel which has been carried out as part of the current research.\cite{VanderPoel2013} Additionally, the simulated flow field in a tube without a jet pump is compared with an analytic wave propagation model\cite{Tijdeman1975} and excellent agreement is obtained.

\subsection{Data analysis\label{sec:data_analysis}}

From the CFD results, a transient solution field for all flow variables is obtained. In order to obtain the complex amplitudes (denoted with the subscript $_1$), a point-wise discrete Fourier term is calculated for the specified wave frequency using data from the last five simulated wave periods. The jet pump velocity amplitude $|u_{1,\mathit{JP}}|$ is calculated using an area-weighted average of the velocity amplitude at the local grid points in the smallest opening of the jet pump (see Fig.~\ref{fig:jetpumpgeom}). The time-averaged variables (denoted with the subscript $_2$) are calculated by averaging the time-series solution over an integer number of wave periods, thus eliminating all first order effects. Note that for the time-averaged streaming velocity field a density-weighted average is applied:
\begin{equation}
\mathbf{u}_2 = \frac{\langle\rho\mathbf{u}\rangle}{\langle\rho\rangle},
\end{equation}
where $\langle\dots\rangle$ indicates time-averaging.

\subsection{Computational mesh\label{sec:mesh}}

\begin{figure}
\centering
\includegraphics[width=.5\textwidth]{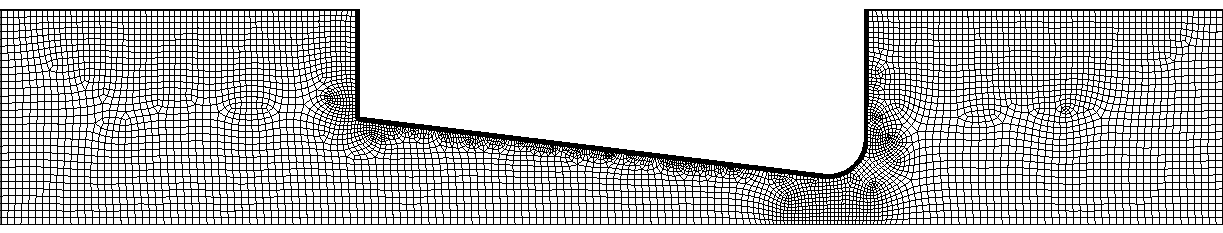}
\caption{Computational mesh, close-up near jet pump.}
\label{fig:mesh}
\end{figure}

The resolution of the computational mesh is defined based on the maximum element size in various regions of the domain. A maximum element size inside the jet pump region of \SI{1}{\mm} is used which is refined up to a maximum size of \SI{0.5}{\mm} near the jet pump waist as is visible in Fig.~\ref{fig:mesh}. Furthermore, the mesh is refined near the viscous boundary layer in the jet pump region such that a minimum of $N_\mathit{el,BL}=10$~elements reside within $1\cdot\delta_\nu$ from the no-slip walls. Here $\delta_\nu = \sqrt{2\mu/\omega\rho}$ is the viscous penetration depth which is $\delta_\nu  = \SI{0.22}{\mm}$ for $f=\SI{100}{\Hz}$. At a distance of \SI{50}{\mm} away from the jet pump, a transition to a structured mesh is applied to allow for uncoupling of the gradients in the $x$ and $r$ directions and consequently for the use of a large aspect ratio as the gradients far away from the jet pump are much larger in the radial direction than they are in the axial direction. The axial element size grows towards the axial boundaries up to a maximum element size of \SI{10}{\mm} which is sufficient for solving the acoustic wave propagation. In the radial direction, a maximum element size of \SI{1}{\mm} is used throughout the outer tube. For the current geometry, this yields a total mesh size of $\SI{36236}{}$~nodes.

In order to validate the computational mesh, the results of three different mesh resolutions are compared based on several key outcome quantities: the dimensionless time-averaged pressure drop ($\Delta p_2^*$, Eq.~\ref{eq:dp2_star}), the dimensionless acoustic power dissipation ($\Delta \dot{E}_2^*$, Eq.~\ref{eq:dE2_star}) and the propagation distance of the vortex street after ten wave periods, $\ell_p$. The specified element sizes for the three different meshes are listed in Table~\ref{tab:meshes}. In Table.~\ref{tab:meshstudy} the total number of nodes is shown. They increase by approximately a factor of two between the subsequent mesh refinements.
\begin{table}
\centering
\caption{Specified element sizes for three different meshes used for validation purposes. For all other results presented, the medium mesh is used.}
\label{tab:meshes}
\begin{tabular}{llp{1.6cm}p{1.6cm}p{1.6cm}}
 &  & \multicolumn{3}{c}{max. element size} \\
 Mesh	 & 		$N_\mathit{el,BL}$  & jet pump region & jet pump waist & outer tube \\
\hline
Coarse 	& 5 & \SI{2}{\mm} & \SI{1}{\mm} & \SI{10}{\mm} \\
Medium 	& 10 & \SI{1}{\mm} & \SI{0.5}{\mm} & \SI{10}{\mm} \\
Fine 	& 20 & \SI{0.5}{\mm} & \SI{0.25}{\mm} & \SI{10}{\mm} \\
\end{tabular}
\end{table}

Table.~\ref{tab:meshstudy} shows the results for the three different meshes using an intermediate wave amplitude ($u_1=\SI{0.8}{\m\per\s}$ which yields $|u_{1,\mathit{JP}}|=\SI{15.3}{\m\per\s}$) and a \SI{7}{\degree} taper angle geometry which is representative for the other simulated cases. The driving frequency is set to \SI{100}{\Hz}. A clear deviation is visible for all outcome quantities between the coarse mesh and the other two meshes while the results between the medium and fine mesh are comparable. The dimensionless pressure drop and acoustic power dissipation obtained with the medium mesh, show a difference of \SI{2.4}{\percent} and \SI{11.1}{\percent} with the fine mesh, respectively. The vortex propagation distance deviates \SI{3.5}{\percent} with respect to the fine mesh. Hence, it was decided to use the medium mesh resolution for all future simulations.
\begin{table}
\centering
\caption{Results of mesh validation study for a jet pump geometry having a taper angle of $\alpha=\SI{7}{\degree}$. The jet pump waist velocity is $|u_{1,\mathit{JP}}|=\SI{15.3}{\m\per\s}$, the driving frequency is $f=\SI{100}{\Hz}$.}
\label{tab:meshstudy}
\begin{tabular}{lllll}
 Mesh	 & 		$N_\mathit{nodes}$  & $\Delta p_2^*$ & $\Delta \dot{E}_2^*$ & $l_p$ \\
\hline
Coarse	& \SI{18442}	& \SI{0.40}	& \SI{0.58}	& \SI{0.81}{\m} \\
Medium	& \SI{36236}	& \SI{0.84}	& \SI{0.86}	& \SI{0.92}{\m} \\
Fine	& \SI{84618}	& \SI{0.86}	& \SI{0.97}	& \SI{0.96}{\m} \\
\end{tabular}
\end{table}

\section{Results and Discussion}

The described computational model has been used to investigate a range of wave amplitudes with the two described jet pump geometries ($\alpha=$~\SIlist{7;15}{\degree}). It will be shown that the jet pump performance can be scaled based on the acoustic displacement amplitude with respect to the jet pump dimensions. Defining the acoustic displacement amplitude in the jet pump waist under the assumption of a sinusoidal jet pump velocity as
\begin{equation}
\xi_{1,\mathit{JP}} = \frac{|u_{1,\mathit{JP}}|}{2 \pi f},
\label{eq:x1JP} 
\end{equation}
the two Keulegan-Carpenter numbers can be defined based on the jet pump length and waist diameter, respectively:
\begin{equation}
KC_L = \frac{\xi_{1,\mathit{JP}}}{L_\mathit{JP}}, \label{eq:KCl}
\end{equation}
\begin{equation}
KC_D = \frac{\xi_{1,\mathit{JP}}}{2R_{s,\mathit{eff}}}. \label{eq:KCd}
\end{equation}
Following Smith and Swift,\cite{Smith2003a} $KC_D$ is similar to the dimensionless stroke length $L_0/h$ while $KC_L$ is one of the suggested additional dimensionless parameters that may affect the results. By investigating two jet pumps of different lengths, it will be shown that $KC_L$ is of high relevance to scale the jet pump performance properly.

Different observed flow regimes will be distinguished and the corresponding flow fields will be described. Typical axial profiles of pressure, velocity and acoustic power will be described and used to define the jet pump performance. Finally, the time-averaged pressure drop and acoustic power dissipation will be scaled and shown as a function of the two Keulegan-Carpenter numbers. In this way, the jet pump performance will be related to the observed flow phenomena and includes the influence of the jet pump taper angle.

\subsection{Flow regimes\label{sec:flow_regimes}}

\begin{figure}
\centering
\subfloat[Oscillatory vortex pair on both sides, no jetting observed. $KC_L=0.09$, $KC_D=0.46$, $\Delta p_2^*=0.04$.\label{fig:flowregime1}]{\includegraphics[width=.7\textwidth]{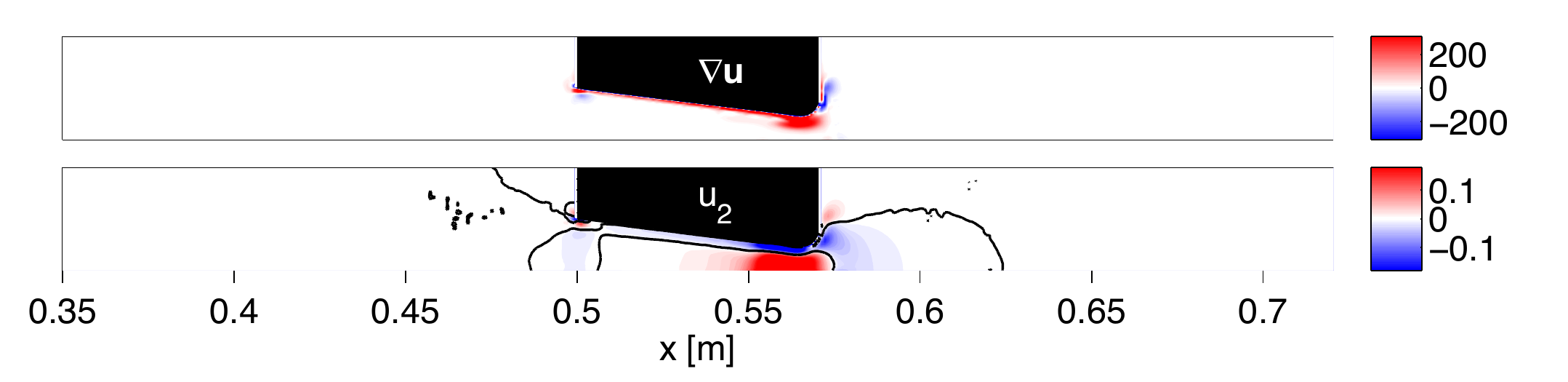}}\\
\subfloat[Propagating vortex to right side, oscillating vortex pair on left side. $KC_L=0.18$, $KC_D=0.92$, $\Delta p_2^*=0.40$.\label{fig:flowregime2}]{\includegraphics[width=.7\textwidth]{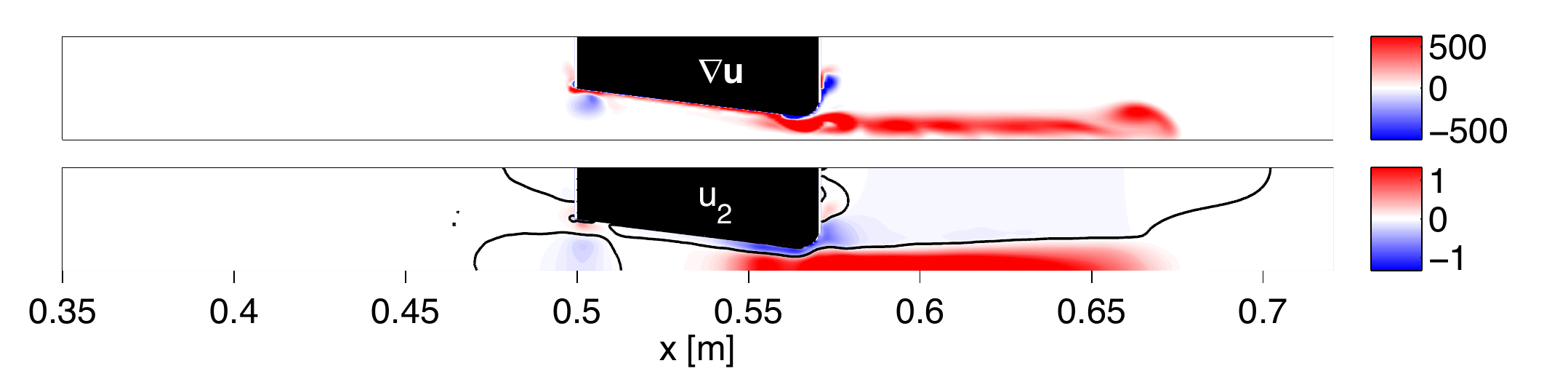}}\\
\subfloat[Propagating vortex on both sides. $KC_L=0.35$, $KC_D=1.74$, $\Delta p_2^*=0.84$.\label{fig:flowregime3}]{\includegraphics[width=.7\textwidth]{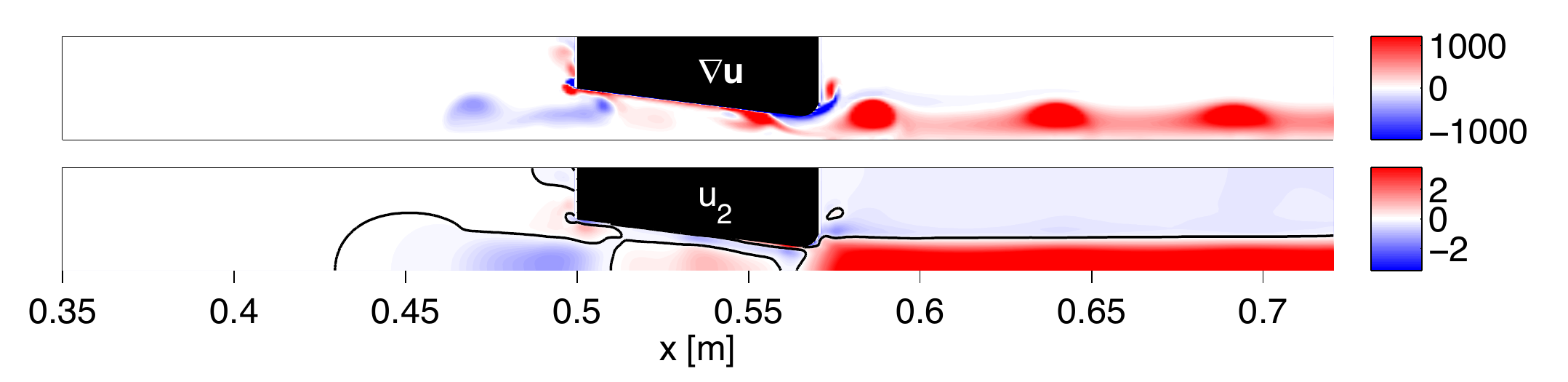}}\\
\subfloat[Left propagating vortex from waist of jet pump, flow separation inside the jet pump occurs. $KC_L=1.17$, $KC_D=5.90$, $\Delta p_2^*=0.46$.\label{fig:flowregime4}]{\includegraphics[width=.7\textwidth]{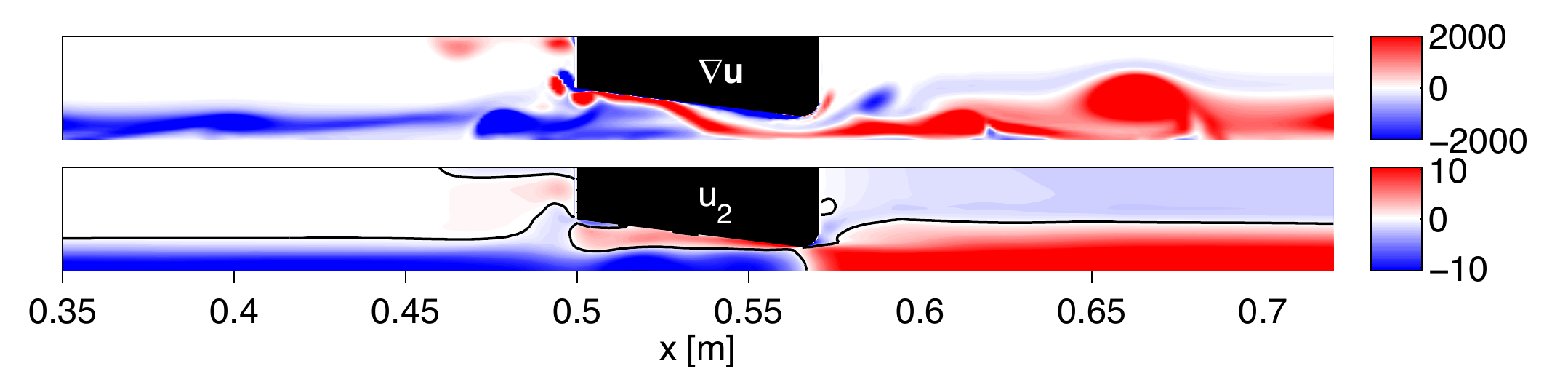}}
\caption{Four different flow regimes are distinguished based on the Keulegan-Carpenter numbers $KC_L$ and $KC_D$ using the instantaneous vorticity fields $\nabla \mathbf{u}$~$[\SI{1}{\per\s}]$ at $t=t_\mathit{max}$ (top) and streaming velocity fields $u_2$~$[\SI{}{\m\per\s}]$ (bottom) around the jet pump for the $\alpha=\SI{7}{\degree}$ geometry, $f=\SI{100}{\Hz}$. Black line in streaming velocity field indicates transition from positive to negative velocity (color online).}
\label{fig:u2+vort}
\end{figure}

\begin{figure}
\centering
\subfloat[\label{fig:uRadHalfJet1}]{\includegraphics[width=0.25\textwidth]{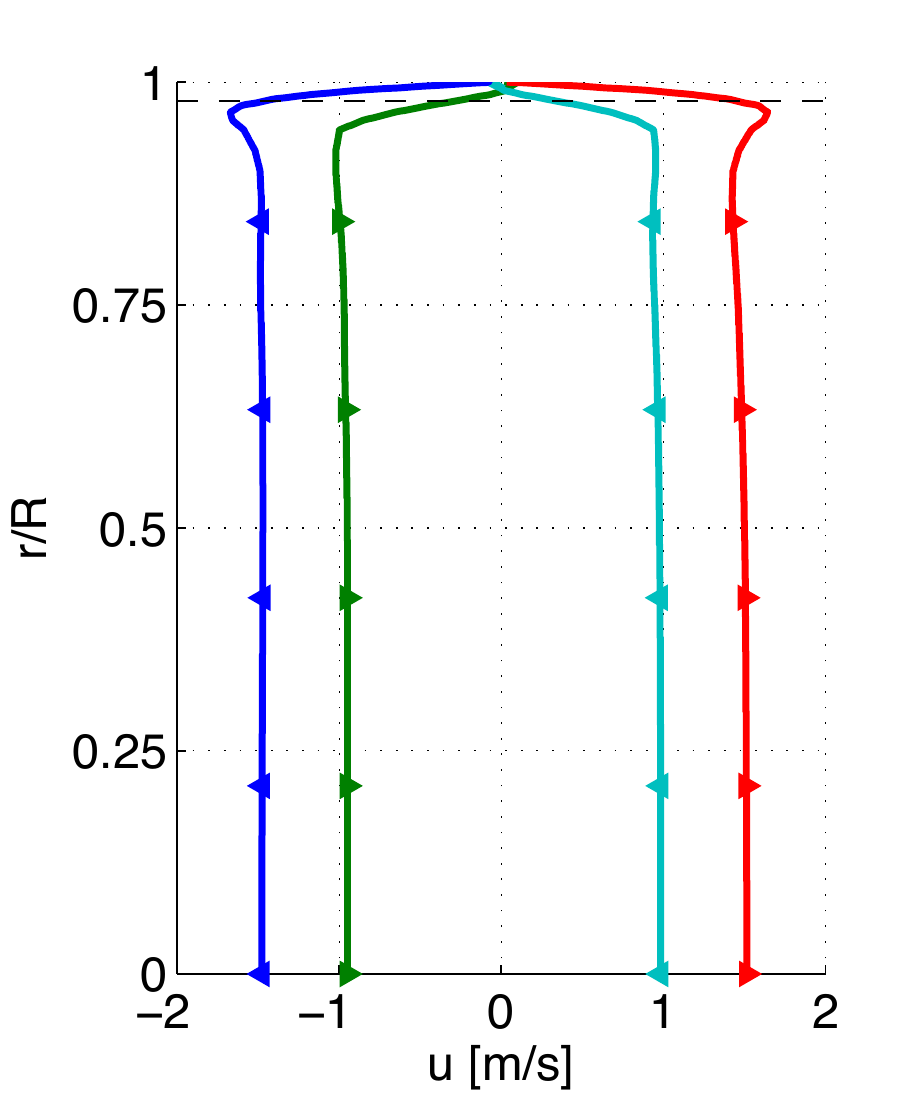}}
\subfloat[\label{fig:uRadHalfJet2}]{\includegraphics[width=0.25\textwidth]{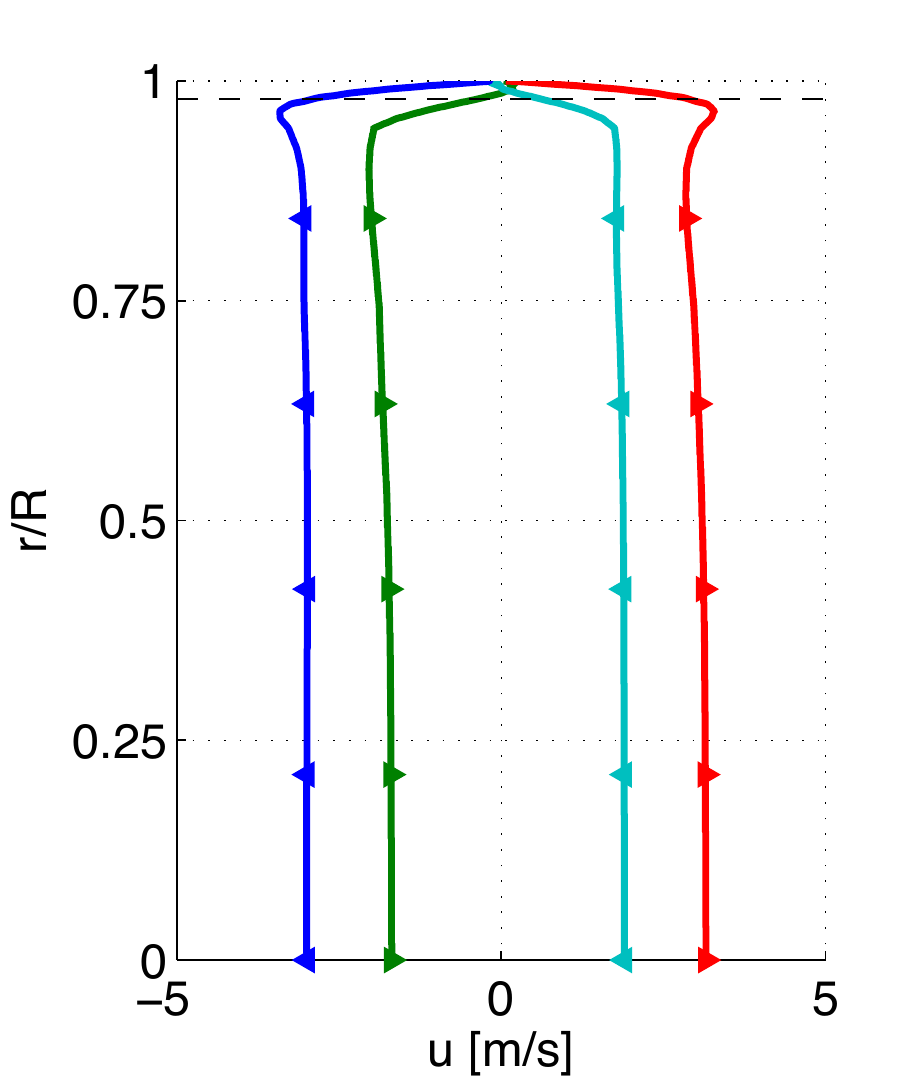}}
\subfloat[\label{fig:uRadHalfJet3}]{\includegraphics[width=0.25\textwidth]{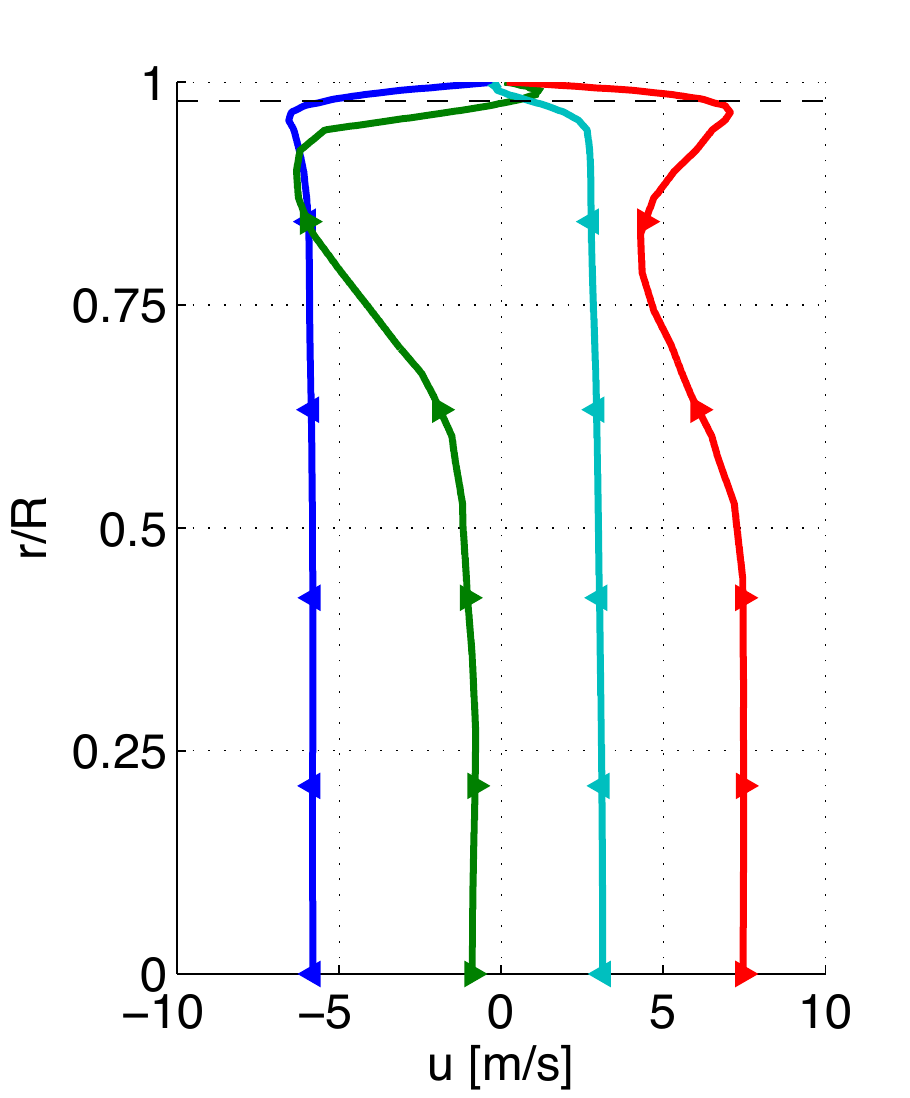}}
\subfloat[\label{fig:uRadHalfJet4}]{\includegraphics[width=0.25\textwidth]{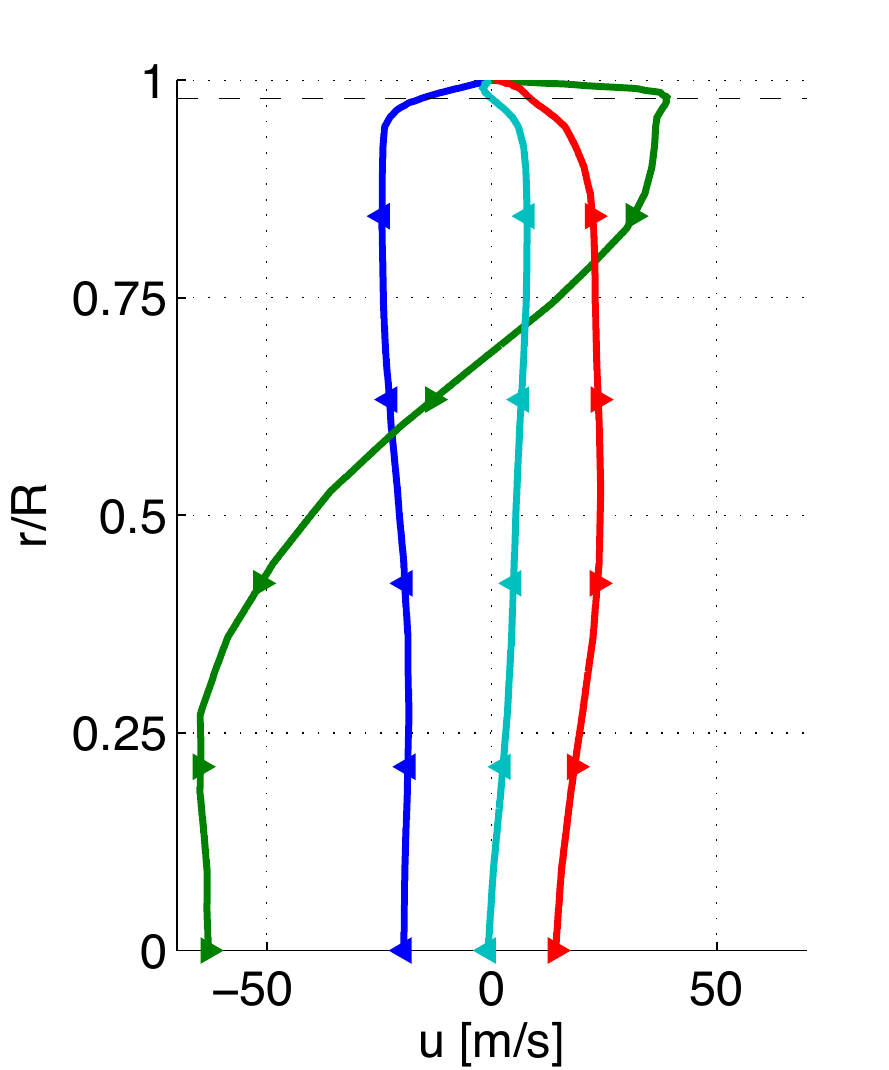}}
\caption{Axial velocity $u$ over the radius at four time instances during the last wave period ($\varphi=$ \SIlist{0;90;180;270}{\degree}) halfway inside the jet pump ($x=L_0+L_\mathit{JP}/2$). Figures correspond to the four different flow regimes distinguished in Fig.~\ref{fig:u2+vort}. Arrows indicate acceleration and deceleration: $\blacktriangleright$ for $\partial u/\partial t>0$ and $\blacktriangleleft$ for $\partial u/\partial t<0$. $r/R=1$ denotes the jet pump wall and $r/R=0$ is the centerline location. The dashed line indicates the thickness of the viscous boundary layer ($\delta_\nu$) at the jet pump wall (color online).}
\label{fig:uRadHalfJet}
\end{figure}

Independent of the jet pump geometry or frequency, four different flow regimes can be distinguished. Examples of these flow regimes are shown in Fig.~\ref{fig:u2+vort} for the \SI{7}{\degree} taper angle jet pump and a driving frequency of $f=\SI{100}{\Hz}$. The top graph of each figure shows the instantaneous vorticity field at the last simulated time-step $t_\mathit{max}=\SI{0.1}{\s}$.  The centers of the propagating vortex rings can be identified as local maxima in the instantaneous vorticity field. The bottom graph of each figure shows the time-averaged velocity field $u_2$. The black line denotes the location of zero streaming velocity. Fig.~\ref{fig:uRadHalfJet} shows the axial velocity over the radius inside the jet pump. The different lines are separated $\varphi=\pi/2$ in time. Each figure represents a different flow regime which corresponds to the flow regimes shown in Fig.~\ref{fig:u2+vort}.

In all simulated cases, a vortex pair is formed on either side of the jet pump. However, for low amplitudes the vortex pairs are not shed and merely oscillate locally with the acoustic field. This results in a zero time-averaged pressure drop and negligible acoustic power dissipation.  An example of this flow regime is shown in Fig.~\ref{fig:flowregime1} where on either side of the jet pump a small vortex can be observed. The corresponding velocity profiles inside the jet pump are shown in Fig.~\ref{fig:uRadHalfJet1}, the profiles are identical but opposite during the forward and backward flow direction representing a pure harmonic oscillation. The influence of the viscous boundary layer is visible but further away from the boundary a constant velocity is observed.

If the displacement amplitude is larger than the radius of one of the jet pump openings, the vortex pair on the corresponding side is shed and propagation starts. Hence, for $KC_D>0.5$ vortex propagation on the right side of the jet pump can be observed. An example of this flow regime is shown in Fig.~\ref{fig:flowregime2} where $KC_D=0.92$. In the streaming velocity field (bottom graph), a steady jet in the positive $x$-direction can be observed. The occurrence of vortex shedding corresponds well to an increase in the time-averaged pressure drop. The velocity profiles inside the jet pump are shown in Fig.~\ref{fig:uRadHalfJet2} and are comparable to the first flow regime where a harmonic oscillation is observed.

Fig.~\ref{fig:flowregime3} shows an intermediate flow regime where vortex propagation to the left side of the jet pump can be observed in addition to the right-sided propagation. However, the flow field is still rather asymmetric on both sides of the jet pump which results in a high time-averaged pressure drop. The vortex propagation speed $u_v$ is strongly dependent on the velocity amplitude in the jet pump waist. Comparing this flow regime to the previous, the wave amplitude and correspondingly the vortex propagation speed has increased. This results in vortices clearly separated from each other. In the streaming velocity field, a recirculation zone inside the jet pump can be observed which is caused by a difference in the velocity profile during the accelerating and decelerating phase. Fig.~\ref{fig:uRadHalfJet3} shows the velocity profile inside the jet pump at four different time instances. When the fluid is accelerating ($\partial u/\partial t>0$, indicated by $\blacktriangleright$), a region near the jet pump wall can be observed where the velocity is lower compared to the bulk velocity, regardless of the direction of the bulk flow. This is initiated when the negative bulk velocity is at its maximum and the acceleration changes sign. During the remainder of the acceleration phase, the velocity near the jet pump wall (but outside the viscous boundary layer) ``lags'' the bulk flow. When the fluid starts decelerating (indicated by $\blacktriangleleft$), a pure acoustic velocity profile is again observed. This difference in velocity profiles leads to a time-averaged recirculation inside the jet pump.

The fourth flow regime is observed when the displacement amplitude is larger than the jet pump length ($KC_L>1)$. An example of this case is shown in Fig.~\ref{fig:flowregime4}. Vortices are now displaced from the right jet pump tip through the jet pump to the left, resulting in an additional steady jet in the negative $x$-direction. During the other half of the acoustic period, vortices shed from the left jet pump tip are displaced through the jet pump to the right and propagate in the positive $x$-direction, contributing to the existing steady jet on this side of the jet pump. These vortex rings are smaller and propagate with a lower speed because they are shed at a location where the velocity amplitude is lower than in the jet pump waist. After some distance (outside the shown region), the smaller rings merge with the larger vortex rings originating from the jet pump waist. 

The existing jet through the jet pump causes time-averaged flow separation inside the jet pump which is visible in the bottom graph of Fig.~\ref{fig:flowregime4}. In contrast with the previous flow regimes, now a positive streaming velocity exists close to the jet pump wall. Examining the instantaneous velocity profiles inside the jet pump in Fig.~\ref{fig:uRadHalfJet4} gives more insight into the flow separation process. When the flow starts accelerating during the backward flow phase, the local shear stress at the jet pump wall, $\left.\mu\frac{\partial u}{\partial r}\right|_{r=R}$, becomes zero and the flow separation is initiated (line marked with $\blacktriangleright$ and negative centerline velocity). After the flow reversal, the flow becomes uni-directional again for the remainder of the wave period. This is independent of the sign of $\partial u/\partial t$. Although the radius of curvature meets the steady flow criterion for a ``smooth'' contraction ($R_c/D_s=0.36 > 0.15$), it is expected that the curvature plays an important role in the flow separation process.\cite{Smith2003b} A detailed investigation of this geometric parameter is part of future work.

\subsection{Axial profiles}

Before describing the relation between the jet pump performance and the jet pump waist velocity, the results for a typical simulation case are described for the \SI{7}{\degree} taper angle jet pump geometry with a driving frequency of $f=\SI{100}{\Hz}$. These results correspond to the flow field shown in Fig.~\ref{fig:flowregime3} where a steady jet to the right side of the jet pump exists and vortex propagation to the left side of the jet pump has just started. On the left boundary condition, a far field velocity amplitude of $u_1=\SI{0.8}{\m\per\s}$ is specified resulting in a velocity amplitude in the jet pump waist of $|u_{1,\mathit{JP}}| = \SI{15.3}{\m\per\s}$.

\begin{figure}
\centering
\subfloat[Velocity amplitude $|u_1|$ at $r=0$ (gray, solid) and area-averaged over the cross-section (gray, dashed). Black lines show streaming velocity $u_2$ at $r=0$ (solid) and at $r=\frac{2}{3}R_0$ (dashed).\label{fig:u1u2_784}]{\includegraphics[width=.45\textwidth]{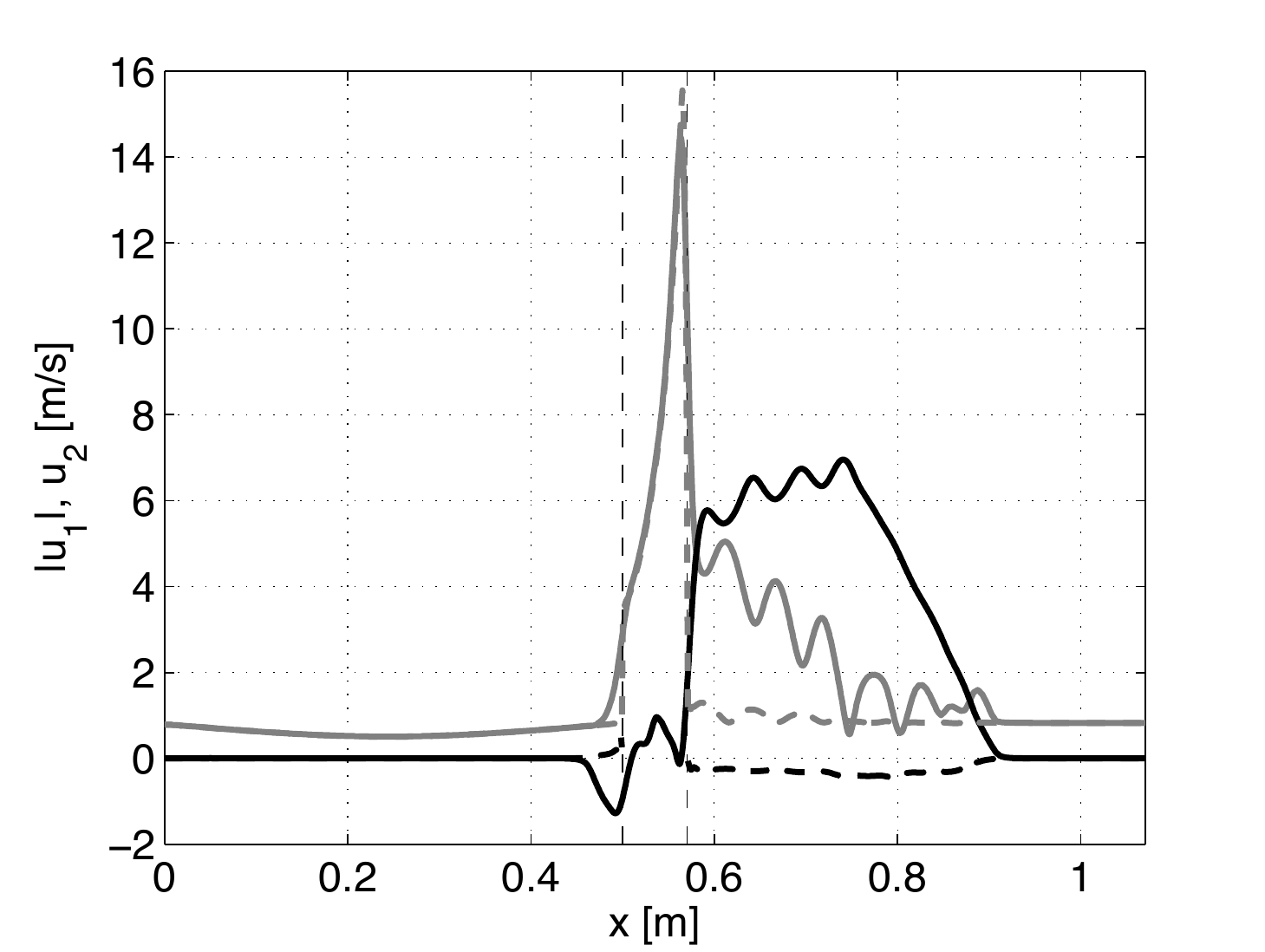}}
\subfloat[Area-averaged pressure amplitude $|p_1|$ (gray, left axis) and time-averaged pressure $p_2$ (black, right axis). \label{fig:p1p2_784}]{\includegraphics[width=.45\textwidth]{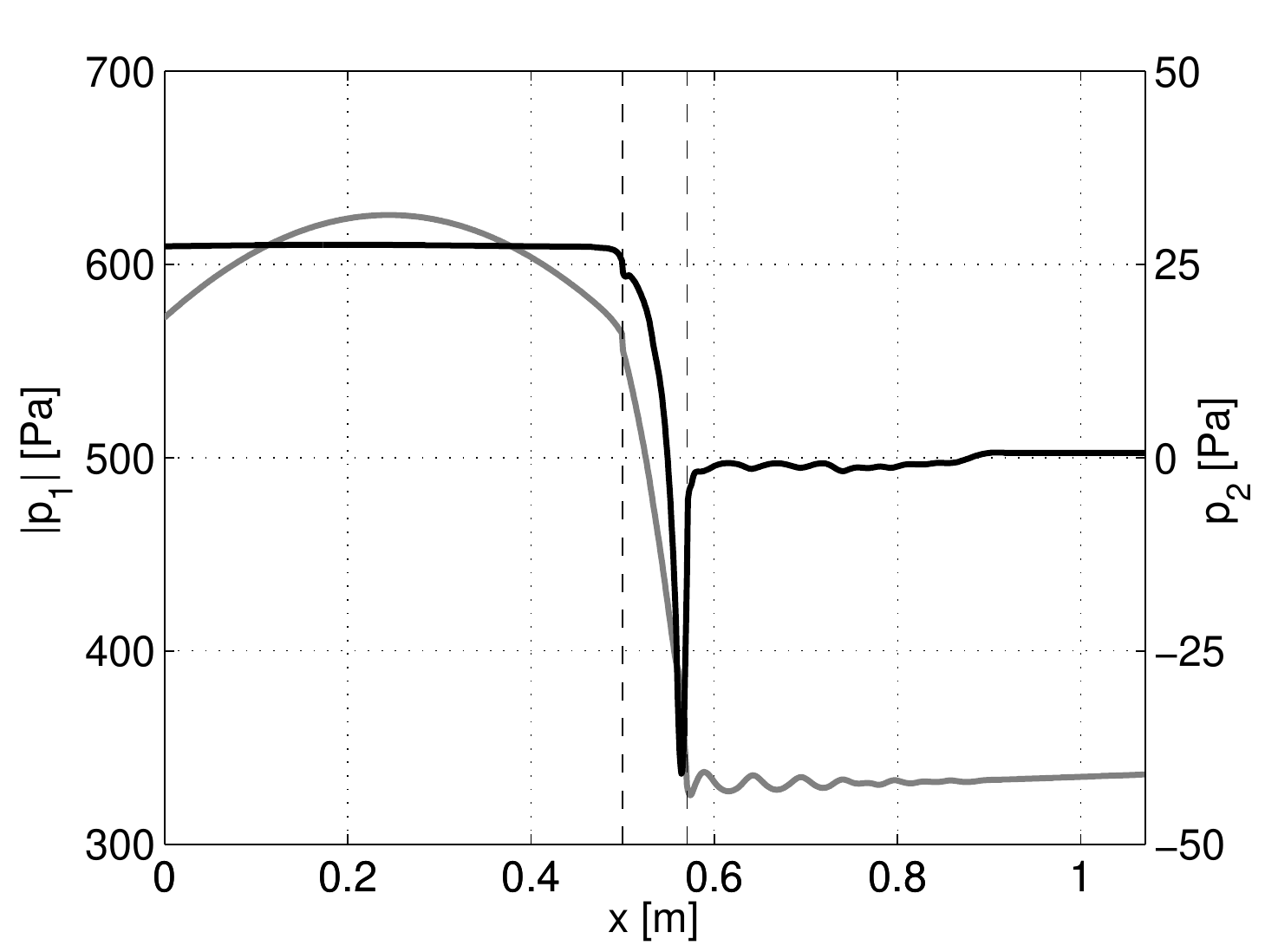}}\\
\subfloat[Acoustic power $\dot{E}_2$.\label{fig:E2_784}]{\includegraphics[width=.45\textwidth]{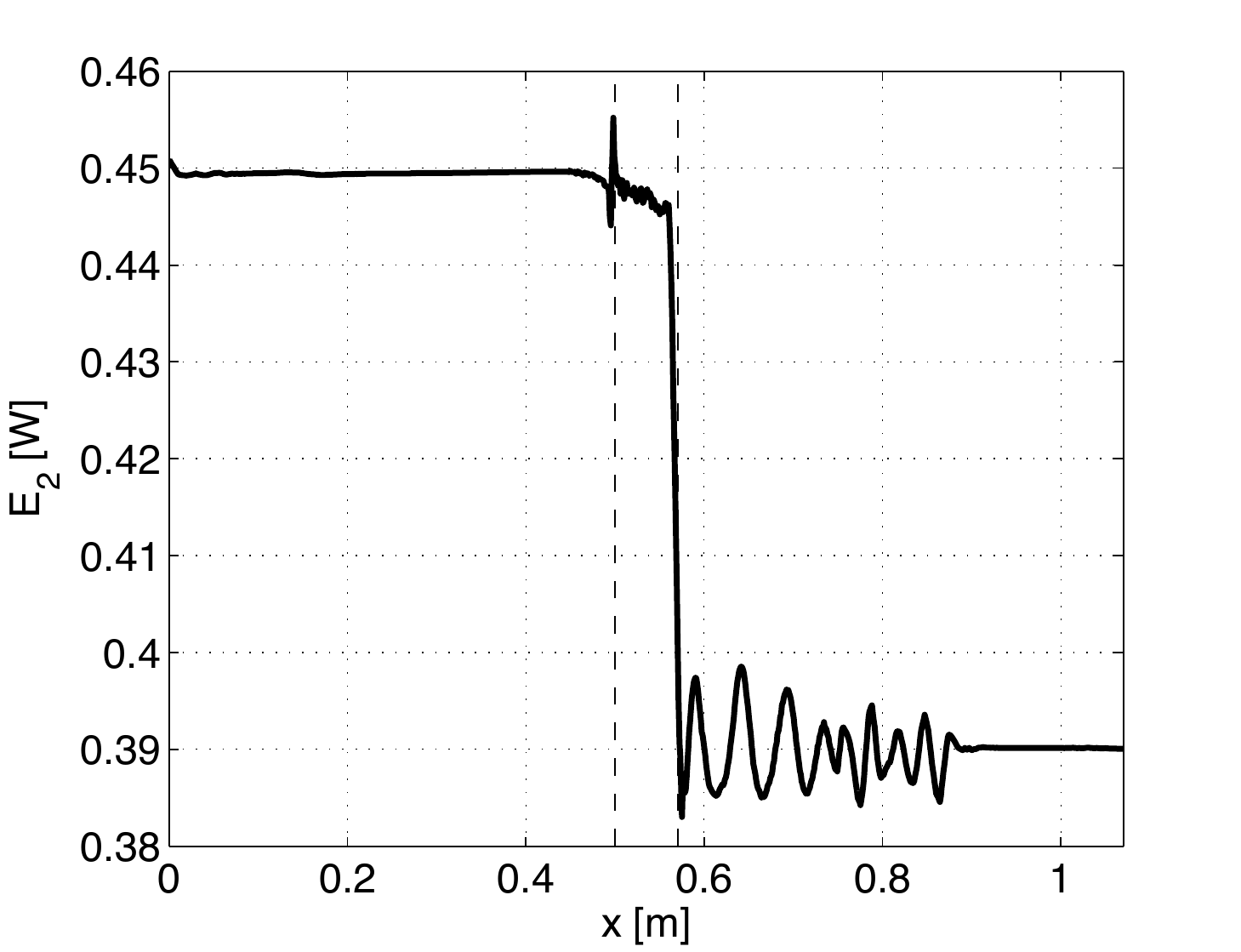}}
\caption{Velocity, pressure, and acoustic power along $x$-axis using a jet pump taper angle of $\alpha=\SI{7}{\degree}$. Jet pump waist velocity amplitude is $|u_{1,\mathit{JP}}|=\SI{15.3}{\m\per\s}$ and $f=\SI{100}{\Hz}$, corresponding to the flow fields in Fig.~\ref{fig:flowregime3}. Vertical dashed lines indicate the exits of the jet pump.}
\label{fig:axial}
\end{figure}

Fig.~\ref{fig:u1u2_784} shows the velocity amplitude and the time-averaged velocity profile along the $x$-axis. The area-averaged velocity amplitude (dashed gray line) shows nearly incompressible behavior where the velocity is inversely proportional to the cross-sectional area. The volume flow rate $U_1$ to the right of the jet pump is constant to within \SI{0.7}{\percent}. The irregularities in the velocity amplitude at the centerline (solid gray line) to the right of the jet pump are caused by the  vortex shedding. A positive time-averaged centerline velocity (solid black line) to the right of the jet pump indicates a steady jet being formed from the jet opening. Note that the jet has only propagated a distance of $\ell_p=\SI{0.35}{\m}$ from the jet pump. Furthermore, the steady jet is balanced by a mean flow in the opposite direction closer to the tube wall (dashed black line) to ensure a zero mean mass flux over the cross-section.  

Fig.~\ref{fig:p1p2_784} shows both the pressure amplitude (gray line, left axis) and the time-averaged pressure (black line, right axis). The constant pressure amplitude to the right of the jet pump indicates that a traveling wave exists in this part of the domain. On the left side of the jet pump a standing wave component is present due to reflection of the acoustic wave off the jet pump surface. This is confirmed by calculating the reflection coefficient. On the right side $|R_R|=\SI{1.16}{\percent}$ while on the left side $|R_L|=\SI{49.9}{\percent}$. The time-averaged pressure in Fig.~\ref{fig:p1p2_784} shows a clear drop between left and right of the jet pump which is one of the main measures of performance of the jet pump. While one could simply subtract the time-averaged pressure $p_2$ at two locations on either side of the jet pump, this would lead to inconsistent results due to the influence of vortex shedding on the time-averaged pressure profile. This is visible in Fig.~\ref{fig:p1p2_784} and becomes more dominant at higher wave amplitudes. Because the vortex propagation speed is much lower than the speed of sound, time-averaging over a multiple of the acoustic period does not remove the contribution of the vortex propagation to the pressure field. Alternatively, a spatially average of $p_2(x)$ on either side of the jet pump is calculated with specific averaging intervals starting at a distance of $2 \cdot \xi_{1,\mathit{JP}}$ from the jet pump up to the total vortex propagation distance $\ell_p$. When no vortex street is present on the corresponding side of the jet pump, the averaging is carried out up to the axial extremities of the domain. The two resulting spatial averages, left and right of the jet pump, are then subtracted yielding the time-averaged pressure drop $\Delta p_2$ across the jet pump. For the case shown in Fig.~\ref{fig:axial}, a total pressure drop of $\Delta p_2=\SI{28.8}{\Pa}$ is generated.

Fig.~\ref{fig:E2_784} shows the axial profile of the acoustic power $\dot{E}_2$. The acoustic power is determined by the pressure and velocity fields and can be calculated either by direct time integration of the transient solution or by using the calculated complex wave amplitudes. In the first method all resolved higher order effects will be included while in the latter only the first order acoustics is taken into account. While the integration method will provide the most complete solution, the method based on first order variables is most relevant for thermoacoustic applications as typically only the power transported by the first harmonic will contribute to a device's efficiency and any conversion to higher harmonics are considered losses.\cite{Swift2002_acousticpower} Using the calculated amplitudes of pressure $p_1$ and volume flow rate $U_1$, the acoustic power is defined as
\begin{equation}
\dot{E}_2(x) = \frac{1}{2}\Re{\left[\tilde{p}_1(x)U_1(x)\right]},
\label{eq:E2}
\end{equation}
where $\tilde{p}_1(x)$ is the complex conjugate of the pressure amplitude area-averaged over the local cross section. The acoustic power dissipation across the jet pump $\Delta\dot{E}_2$ is determined in a similar manner as the time-averaged pressure drop. For the current case this yields $\Delta\dot{E}_2=\SI{58.8}{\milli\watt}$. 

\begin{figure}
\centering
\subfloat[Time-averaged pressure drop $\Delta p_2$.\label{fig:dp2}]{\includegraphics[width=.45\textwidth]{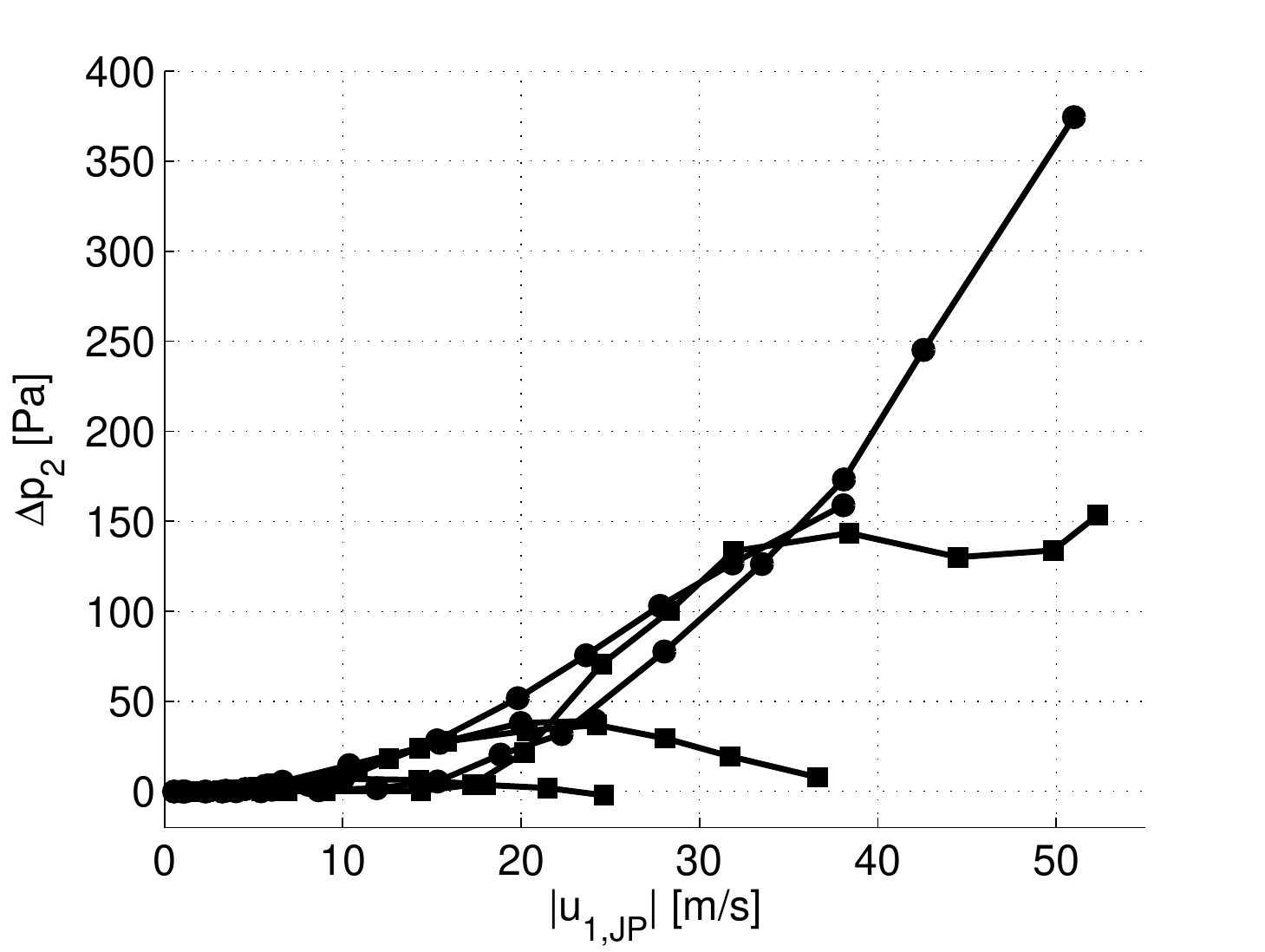}}
\subfloat[Acoustic power dissipation $\Delta \dot{E}_2$.\label{fig:dE2}]{\includegraphics[width=.45\textwidth]{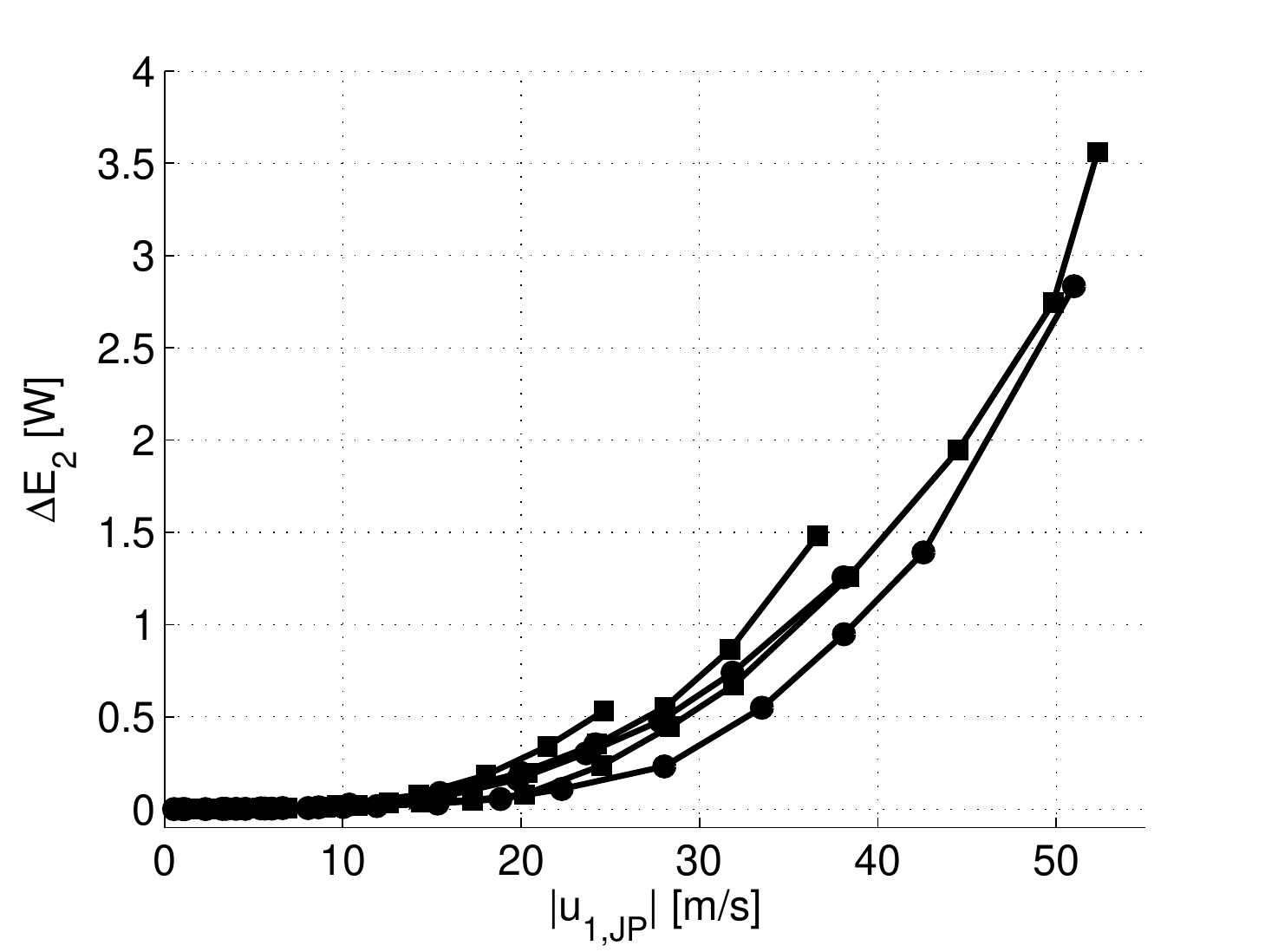}}
\caption{Time-averaged pressure drop and acoustic power dissipation for two different jet pump geometries as a function of jet pump waist velocity amplitude, $\alpha=\SI{7}{\degree}$~($\CIRCLE$) and $\alpha=\SI{15}{\degree}$~($\blacksquare$). Three different frequencies included: \SIlist{50;100;200}{\Hz}.}
\label{fig:dp2_dE2}
\end{figure}%

\subsection{Jet pump performance\label{sec:jet_performance}}

By varying the far field velocity amplitude, the relation between the jet pump performance and waist velocity is studied. The jet pump waist velocity amplitude ranges from \SIrange{0.5}{52}{\m\per\s} which corresponds to acoustic displacement amplitudes from \SIrange{0.4}{78}{\milli\m}. From the velocity amplitude, the acoustic Reynolds number is defined as
\begin{equation}
Re = \frac{|u_1|\delta_\nu\rho_0}{\mu_0}.
\label{eq:re}
\end{equation}
Ohmi and Iguchi derived a critical Reynolds number for the transition to turbulence in oscillating pipe flow which can be rewritten to\cite{Ohmi1982,Antao2013}
\begin{equation}
Re_c = 305\left(\frac{D}{\delta_\nu}\right)^{\frac{1}{7}}.
\label{eq:re_c}
\end{equation}
The maximum acoustic Reynolds number in the computational domain occurs at the location of the jet pump waist. For the simulated cases presented in the following, this maximum value falls below the critical Reynolds number which verifies the laminar assumption. Table~\ref{tab:reynolds} shows an overview of the investigated Reynolds numbers and the corresponding critical values.
\begin{table}
\centering
\caption{Acoustic Reynolds numbers in the jet pump waist and critical Reynolds numbers for the simulated cases.}
\label{tab:reynolds}
\begin{tabular}{llll}
 $\alpha$	 & 	$f$ & $\max(Re)$ & $Re_c$ \\
\hline
\SI{7}{\degree}		& \SI{50}{\Hz}	& 486	& 524 \\
\SI{15}{\degree}	& \SI{50}{\Hz}	& 496	& 524 \\
\SI{7}{\degree}		& \SI{100}{\Hz}	& 542	& 551 \\
\SI{15}{\degree}	& \SI{100}{\Hz}	& 521	& 551 \\
\SI{7}{\degree}		& \SI{200}{\Hz}	& 513	& 579 \\
\SI{15}{\degree}	& \SI{200}{\Hz}	& 526	& 579 \\
\end{tabular}
\end{table}

The time-averaged pressure drop and acoustic power dissipation for the two different jet pump taper angles and three different frequencies are shown in Fig.~\ref{fig:dp2_dE2}. The acoustic power dissipation is similar for both geometries and increases approximately with the cube of $|u_{1,\mathit{JP}}|$ as was predicted based on the quasi-steady model (Eq.~\ref{eq:backhaus_dE}). However, for the time-averaged pressure drop a large difference can be observed which is a consequence of the change in taper angle or frequency. Where the quasi-steady model predicts a quadratic relation between $\Delta p_2$ and $|u_{1,\mathit{JP}}|$ (Eq.~\ref{eq:backhaus}), this is clearly not the case over the entire range of velocity amplitude for the simulations. For low amplitudes, a nearly quadratic increase is observed for both geometries. However, for the \SI{15}{\degree} taper angle the pressure drop stagnates and eventually decreases at higher velocity amplitudes. Also, the \SI{7}{\degree} taper angle shows a deviation from the theoretical pressure drop profile and stagnates, but this happens at much higher velocity amplitudes compared to the \SI{15}{\degree} geometry. 

The results at \SI{100}{\Hz} have been compared against preliminary experimental results that have been achieved using identical jet pump geometries in a traveling wave experimental setup at a driving frequency of \SI{113}{\Hz}. The measured time-averaged pressure drop shows the same behavior as the simulation results for both jet pump geometries. The maximum deviation is less than \SI{20}{\percent} in the region where the measured time-averaged pressure drop is significant (for $\Delta p_2>\SI{10}{\Pa}$). The experimental results are part of future research and will be presented in another publication.

The effect of the jet pump geometry and frequency on the time-averaged pressure drop can be explained by scaling the velocity amplitude using the Keulegan-Carpenter numbers based on either the jet pump length or the jet pump waist diameter (Eq.~\ref{eq:KCl} and Eq.~\ref{eq:KCd}, respectively). Moreover, the pressure drop and acoustic power dissipation are scaled according to:\cite{Smith2003a}
\begin{gather}
\Delta p_2^* = \frac{8\Delta p_2}{\rho_0 |u_{1,\mathit{JP}}|^2} ,
\label{eq:dp2_star}\\
\Delta \dot{E}_2^* = \frac{3\pi\Delta \dot{E}_2}{\rho_0\pi R_\mathit{s,eff}^2 |u_{1,\mathit{JP}}|^3},
\label{eq:dE2_star}
\end{gather}
where $\Delta p_2^*$ would represent the difference in minor loss coefficients between the two flow directions assuming the quasi-steady theory (Eq.~\ref{eq:backhaus}) and $\Delta\dot{E}_2^*$ would represent the summation of the minor loss coefficients assuming Eq.~\ref{eq:backhaus_dE} to be valid. 

\begin{figure}
\centering
\includegraphics[width=.45\textwidth]{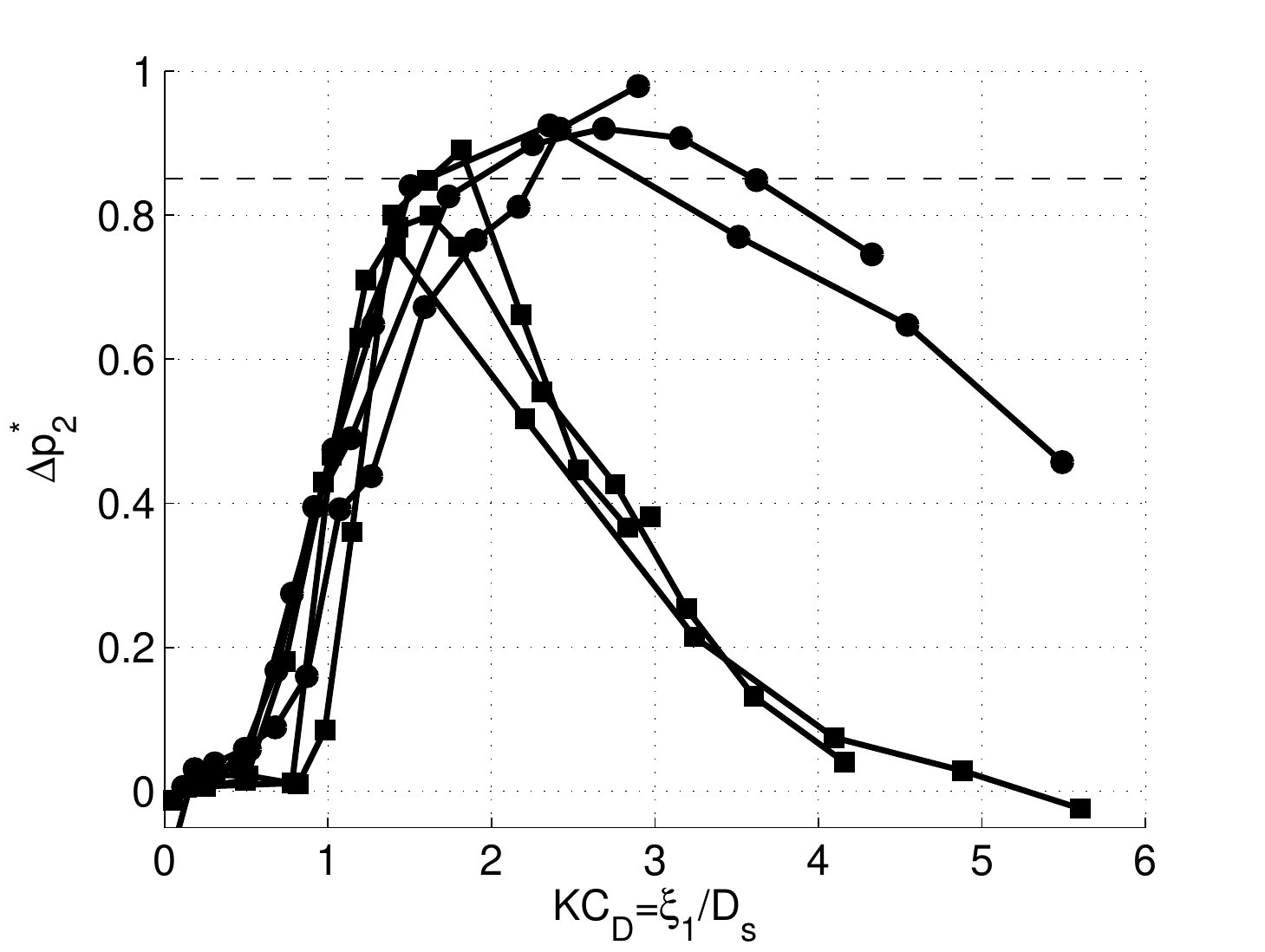}
\caption{Dimensionless pressure drop $\Delta p_2^*$ as a function of $KC_D$ for two different jet pump geometries: $\alpha=\SI{7}{\degree}$~($\CIRCLE$) and  $\alpha=\SI{15}{\degree}$~($\blacksquare$) at \SIlist{50;100;200}{\Hz}. Dashed line indicates quasi-steady approximation.}
\label{fig:dp2_star_KCd}
\end{figure}

\begin{figure}
\centering
\subfloat[Dimensionless pressure drop $\Delta p_2^*$.\label{fig:dp2_star}]{\includegraphics[width=.45\textwidth]{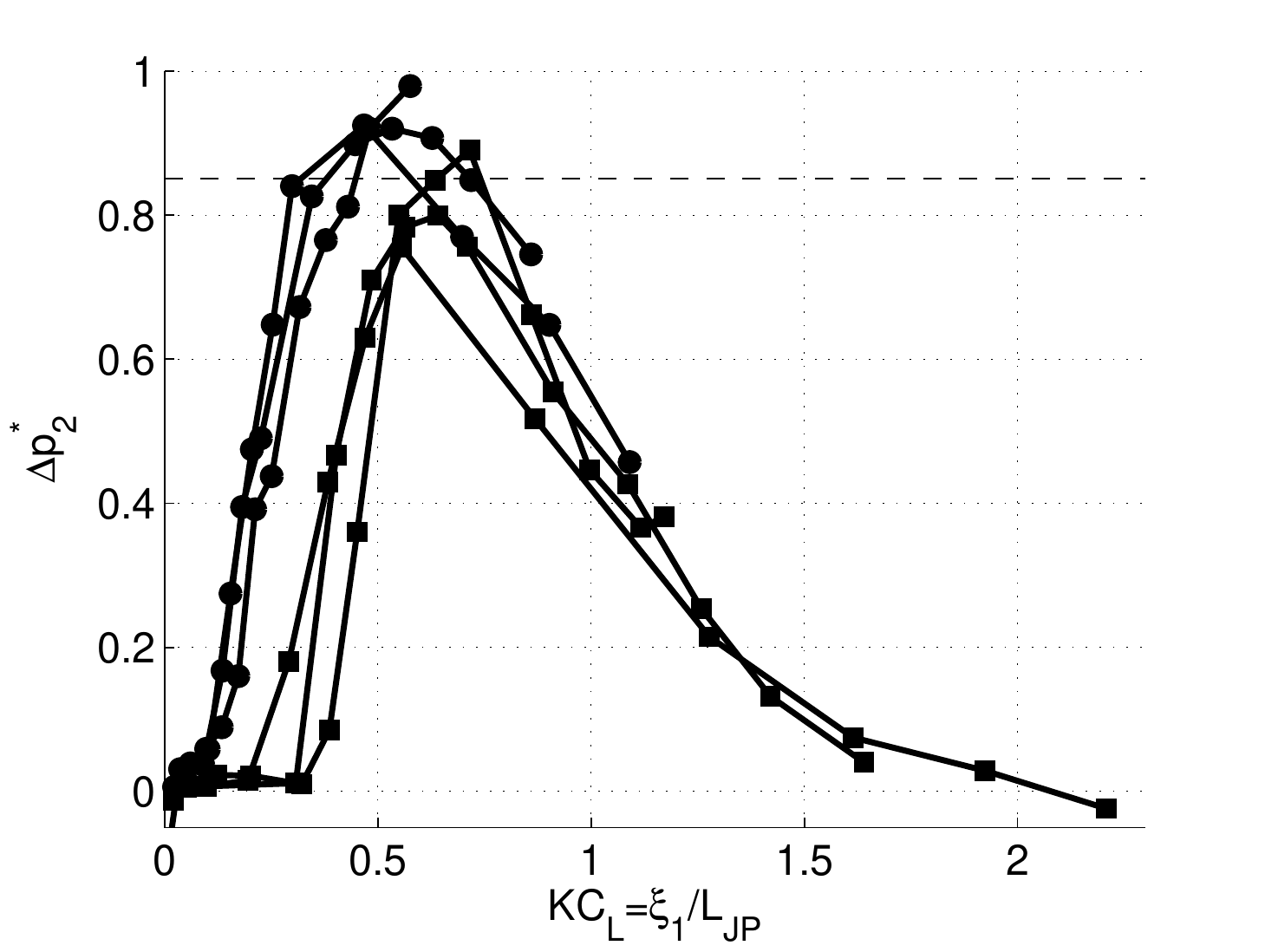}}
\subfloat[Dimensionless acoustic power dissipation $\Delta \dot{E}_2^*$.\label{fig:dE2_star}]{\includegraphics[width=.45\textwidth]{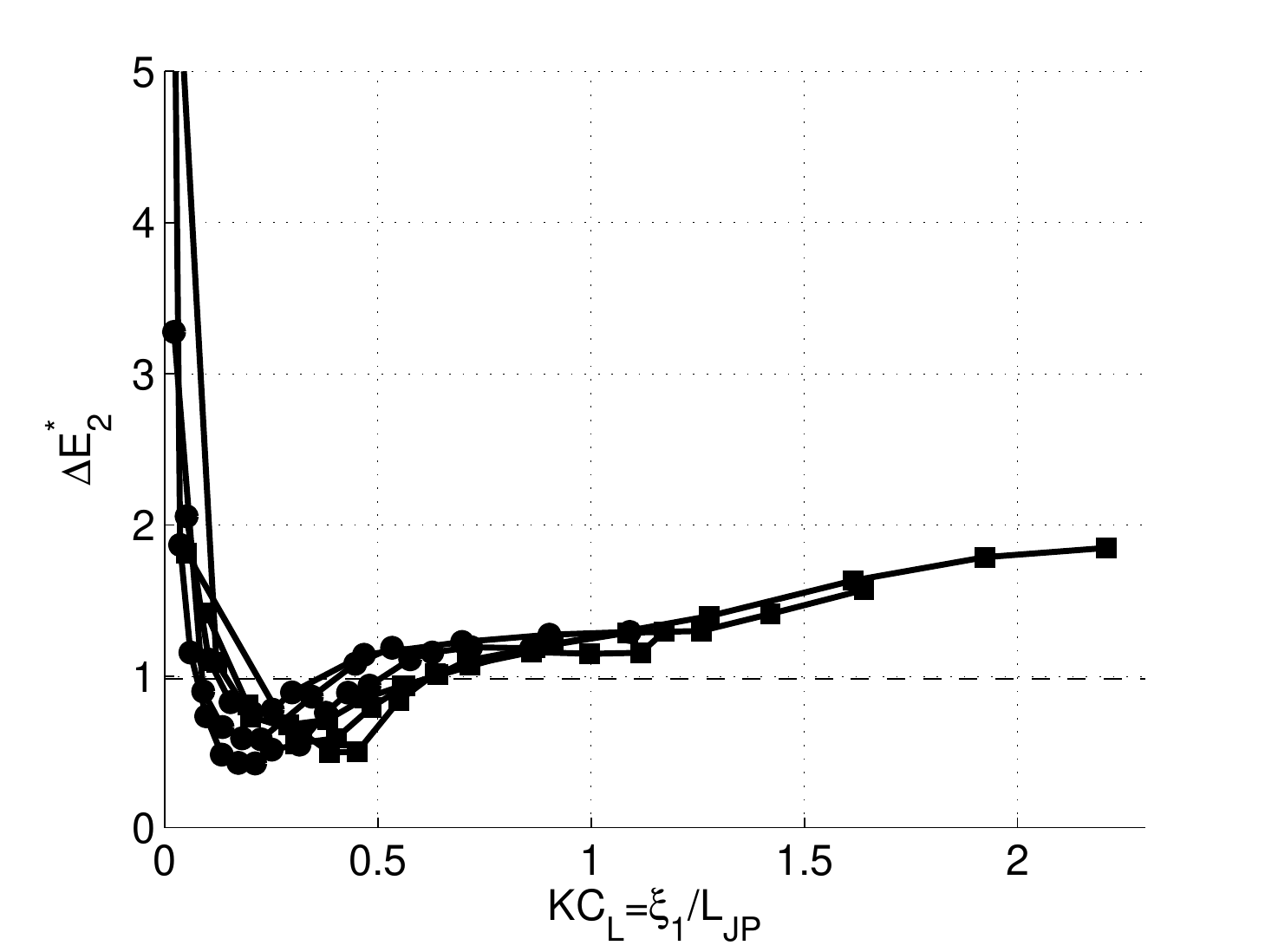}}
\caption{Dimensionless pressure drop and acoustic power dissipation as a function of $KC_L$ for two different jet pump geometries: $\alpha=\SI{7}{\degree}$~($\CIRCLE$) and  $\alpha=\SI{15}{\degree}$~($\blacksquare$) at \SIlist{50;100;200}{\Hz}. Dashed lines indicates quasi-steady approximation.}
\label{fig:dp2_dE2_star}
\end{figure}

The onset of the jet pump working is found to be determined by the Keulegan-Carpenter number based on the jet pump waist diameter, $KC_D$, and is shown in Fig.~\ref{fig:dp2_star_KCd} for the two investigated taper angles and three different driving frequencies. No time-averaged pressure drop is measured for low values ($KC_D<0.5$). This corresponds to the flow regime described in Fig.~\ref{fig:flowregime1} where there are no minor loss inducing flow phenomena observed. In the low amplitude regime, the effect of the jet pump taper angle is negligible and both geometries follow the same line. However, at higher amplitudes the curves become deviant and the effect of jet pump taper angle becomes apparent which is well accounted for by using $KC_L$ rather than $KC_D$. Note that the effect of frequency is well accounted for by using the acoustic displacement amplitude as a scaling parameter rather than the velocity amplitude.

Fig.~\ref{fig:dp2_dE2_star} shows the dimensionless quantities $\Delta p_2^*$ and $\Delta\dot{E}_2^*$ as a function of $KC_L$. Both jet pump taper angles show the same trend indicating that the jet pump performance is directly related to the jet pump length. A maximum in the dimensionless pressure drop is visible around $KC_L\approx 0.5$ for both geometries. As soon as the displacement amplitude becomes large with respect to the jet pump length, the total minor loss coefficient begins to decrease again. This corresponds well to the observed asymmetry in the flow fields as discussed in Section~\ref{sec:flow_regimes}. At $KC_L\approx 0.5$ the flow field is highly asymmetric between both sides of the jet pump (Fig.\ref{fig:flowregime3}) while for $KC_L>1$ a large part of the asymmetry is lost and flow separation occurs. The time-averaged separated flow reduces the ``effective'' diameter of the big opening (Fig.~\ref{fig:flowregime4}). This will consequently lead to more symmetric minor losses and hence to a decrease in the time-averaged pressure drop while the acoustic power dissipation will still increase. In this regime, the quasi-steady approximation is not valid anymore and an adjustment is required.

Comparing the obtained maximum value of $\Delta p_2^*$ with the quasi-steady approximation (dashed line) in Fig.~\ref{fig:dp2_star}, a close match is observed for both taper angles. This suggests the quasi-steady approximation can be used to predict the time-averaged pressure drop but only for the optimal situation where $KC_L\approx 0.5$. The acoustic power dissipation is predicted well for all cases by the quasi-steady model as shown in Fig.~\ref{fig:dE2_star}. The assumed cubic relation between $\Delta\dot{E}_2$ and $|u_{1,\mathit{JP}}|$ is confirmed by the fairly constant value of $\Delta\dot{E}_2^*$, especially for $0.5<KC_L<1.5$. In this region, the jet pump acts as a pure acoustic resistance and its behavior is not affected by a change in flow regime.

\section{Conclusions}

A computational fluid dynamics model is successfully used to simulate the oscillatory flow through two different jet pump geometries under traveling wave conditions. The relation between the obtained time-averaged pressure drop and the acoustic displacement amplitude is investigated together with the acoustic power dissipation. 

Four different flow regimes are distinguished based on the Keulegan-Carpenter numbers and the observed flow phenomena are related to the jet pump performance. For $KC_D<0.5$ no vortex shedding is observed, resulting in a negligible pressure drop. At $KC_L \approx 0.5$, the highly asymmetric flow field leads to a substantial pressure drop and a steady jet to the right side of the jet pump is observed. The measured jet pump performance in this regime corresponds well with the quasi-steady approximation. 

When $KC_L>1$, vortices are shed through the jet pump to the left resulting in an additional jet to the left side of the jet pump and flow separation in the jet pump is observed. Reducing the asymmetry of the flow field consequently leads to a decay in the time-averaged pressure drop but increases the acoustic power dissipation. In this flow regime, an adjustment to the quasi-steady approximation is required to reliably predict the jet pump performance.

Several additional geometric parameters have not been considered in this paper and additional research is required in order to distinguish  more precisely between the different flow phenomena. This should provide insight into whether the decay in $\Delta p_2^*$ for high values of $KC_L$ is solely caused by the jet pump length or by the jet pump taper angle as was postulated in previous work.\cite{Oosterhuis2014} Moreover, the influence of the radius of curvature on the jet pump performance is not investigated here. The radius of curvature is expected to have an effect on the flow separation inside the jet pump and consequently on the overall performance.

A first step towards a better understanding of the physics behind jet pumps is made but more attention is yet required to reliably predict a jet pump's performance. A thorough numerical parameter study on the various (two-dimensional) geometric parameters together with experimental research on three-dimensional geometry variations are the next steps to be undertaken to provide insight into to the jet pump scaling problem.

\medskip

\noindent \textbf{Acknowledgements}

\setlength{\parindent}{0.7cm} 

This work is financially supported by Agentschap~NL as part of the EOS-KTO research program under project number KTOT03009. We would like to thank Bart van der Poel for his work on the time-domain impedance boundary condition which has been carried out as part of a master thesis in Mechanical Engineering at the University of Twente.


\medskip

\begin{thebibliography}{10}
\newcommand{\enquote}[1]{``#1''}
\expandafter\ifx\csname url\endcsname\relax
  \def\url#1{\texttt{#1}}\fi
\expandafter\ifx\csname urlprefix\endcsname\relax\def\urlprefix{URL }\fi
\providecommand{\bibinfo}[2]{#2}
\providecommand{\noopsort}[1]{}
\providecommand{\switchargs}[2]{#2#1}

\bibitem{Backhaus1999}
\bibinfo{author}{S.~Backhaus} and \bibinfo{author}{G.~Swift},
	  \enquote{\bibinfo{title}{{A thermoacoustic Stirling heat engine}}},
  \bibinfo{journal}{Nature} \textbf{\bibinfo{volume}{399}},
  \bibinfo{pages}{335--338} (\bibinfo{year}{1999}).

\bibitem{Gedeon1997}
\bibinfo{author}{D.~Gedeon}, \enquote{\bibinfo{title}{{DC gas flows in Stirling
  and pulse-tube cryocoolers}}}, \bibinfo{journal}{Cryocoolers}
  \textbf{\bibinfo{volume}{9}}, \bibinfo{pages}{385--392}
  (\bibinfo{year}{1997}).

\bibitem{Swift1999}
\bibinfo{author}{G.~Swift}, \bibinfo{author}{D.~Gardner}, and
  \bibinfo{author}{S.~Backhaus}, \enquote{\bibinfo{title}{{Acoustic recovery of
  lost power in pulse tube refrigerators}}}, \bibinfo{journal}{J. Acoust. Soc. Am.} \textbf{\bibinfo{volume}{105}},
  \bibinfo{pages}{711--724} (\bibinfo{year}{1999}).

\bibitem{Backhaus2000}
\bibinfo{author}{S.~Backhaus} and \bibinfo{author}{G.~Swift},
  \enquote{\bibinfo{title}{{A thermoacoustic-Stirling heat engine: detailed
  study}}}, \bibinfo{journal}{J. Acoust. Soc. Am.}
  \textbf{\bibinfo{volume}{107}}, \bibinfo{pages}{3148--3166}
  (\bibinfo{year}{2000}).

\bibitem{Iguchi1982_u-shaped}
\bibinfo{author}{M.~Iguchi}, \bibinfo{author}{M.~Ohmi}, and
  \bibinfo{author}{K.~Meagawa}, \enquote{\bibinfo{title}{{Analysis of free
  oscillating flow in a U-shaped tube}}}, \bibinfo{journal}{Bull. JSME} \textbf{\bibinfo{volume}{25}}, \bibinfo{pages}{1398--1405}
  (\bibinfo{year}{1982}).

\bibitem{Idelchik2007}
\bibinfo{author}{I.~Idel'chik}, \enquote{\bibinfo{title}{{Resistance to flow
  through orifices with sudden change in velocity and flow area}}}, in
  \emph{\bibinfo{booktitle}{Handbook of Hydraulic Resistance}}, edited by
  \bibinfo{editor}{A.~Ginevskiy} and \bibinfo{editor}{A.~Kolesnikov},
  \bibinfo{edition}{4th} edition, chapter~\bibinfo{chapter}{4},
  \bibinfo{pages}{223--275} (\bibinfo{publisher}{Begell House},
  \bibinfo{address}{New York}) (\bibinfo{year}{2007}).

\bibitem{Wakeland2002}
\bibinfo{author}{R.~S. Wakeland} and \bibinfo{author}{R.~M. Keolian},
  \enquote{\bibinfo{title}{{Influence of velocity profile nonuniformity on
  minor losses for flow exiting thermoacoustic heat exchangers (L)}}},
  \bibinfo{journal}{J. Acoust. Soc. Am.}
  \textbf{\bibinfo{volume}{112}}, \bibinfo{pages}{1249--1252} (\bibinfo{year}{2002}).

\bibitem{Petculescu2003}
\bibinfo{author}{A.~Petculescu} and \bibinfo{author}{L.~A. Wilen},
  \enquote{\bibinfo{title}{{Oscillatory flow in jet pumps: nonlinear effects
  and minor losses}}}, \bibinfo{journal}{J. Acoust. Soc. Am.} \textbf{\bibinfo{volume}{113}}, \bibinfo{pages}{1282--1292}
  (\bibinfo{year}{2003}).

\bibitem{Oosterhuis2014}
\bibinfo{author}{J.~P. Oosterhuis}, \bibinfo{author}{S.~B\"{u}hler},
  \bibinfo{author}{D.~Wilcox}, and \bibinfo{author}{T.~H. {Van der Meer}},
  \enquote{\bibinfo{title}{{Computational fluid dynamics analysis of the
  oscillatory flow in a jet pump: the influence of taper angle}}}, in
  \emph{\bibinfo{booktitle}{9th PAMIR International Conference}},
  \bibinfo{pages}{391--395} (\bibinfo{address}{Riga}) (\bibinfo{year}{2014}).

\bibitem{Smith2003a}
\bibinfo{author}{B.~L. Smith} and \bibinfo{author}{G.~W. Swift},
  \enquote{\bibinfo{title}{{Power dissipation and time-averaged pressure in
  oscillating flow through a sudden area change}}}, \bibinfo{journal}{J. Acoust. Soc. Am.} \textbf{\bibinfo{volume}{113}},
  \bibinfo{pages}{2455--2463} (\bibinfo{year}{2003}).

\bibitem{Smith2003b}
\bibinfo{author}{B.~L. Smith} and \bibinfo{author}{G.~W. Swift},
  \enquote{\bibinfo{title}{{A comparison between synthetic jets and continuous
  jets}}}, \bibinfo{journal}{Exp. Fluids}
  \textbf{\bibinfo{volume}{34}}, \bibinfo{pages}{467--472}
  (\bibinfo{year}{2003}).
  
\bibitem{Morris2001}
\bibinfo{author}{P.~Morris}, \bibinfo{author}{S.~Boluriaan}, and
  \bibinfo{author}{C.~Shieh}, \enquote{\bibinfo{title}{{Computational
  thermoacoustic simulation of minor losses through a sudden contraction and
  expansion}}}, in \emph{\bibinfo{booktitle}{7th AIAA/CEAS Aeroacoustics
  \ldots}}, \bibinfo{pages}{1--11}, (\bibinfo{year}{2001}).

\bibitem{Morris2004}
\bibinfo{author}{P.~Morris}, \bibinfo{author}{S.~Boluriaan}, and
  \bibinfo{author}{C.~Shieh}, \enquote{\bibinfo{title}{{Numerical simulation of
  minor losses due to a sudden contraction and expansion in high amplitude
  acoustic resonators}}}, \bibinfo{journal}{Acta Acust. united Ac.}
  \textbf{\bibinfo{volume}{90}}, \bibinfo{pages}{393--409}
  (\bibinfo{year}{2004}).

\bibitem{Boluriaan2003a}
\bibinfo{author}{S.~Boluriaan} and \bibinfo{author}{P.~J. Morris},
  \enquote{\bibinfo{title}{{Suppression of traveling wave streaming using a jet
  pump}}}, in \emph{\bibinfo{booktitle}{41st AIAA Aerospace Sciences Meeting \&
  Exhibit}} (\bibinfo{publisher}{American Institute of Aeronautics and
  Astronautics}, \bibinfo{address}{Reno, NY}), \bibinfo{pages}{1--11}, (\bibinfo{year}{2003}).

\bibitem{ANSYS2011}
\bibinfo{author}{ANSYS}, \enquote{\bibinfo{title}{{ANSYS CFX, Release 14.5}}},
  (\bibinfo{year}{2011}).
  
\bibitem{Aben2010}
\bibinfo{author}{P.~C.~H. Aben},
  \enquote{\bibinfo{title}{{High-amplitude thermoacoustic flow interacting with
  solid boundaries}}}, \bibinfo{type}{PhD thesis}, \bibinfo{school}{Technische
  Universiteit Eindhoven}, (\bibinfo{year}{2010}).

\bibitem{Lycklama2005}
\bibinfo{author}{J.~A. Lycklama \`a Nijeholt}, \bibinfo{author}{M.~E.~H. Tijani}, and \bibinfo{author}{S.~Spoelstra}, \enquote{\bibinfo{title}{{Simulation of a traveling-wave thermoacoustic engine using computational fluid dynamics}}}, \bibinfo{journal}{J. Acoust. Soc. Am.} {\textbf{\bibinfo{volume}{118}}}, \bibinfo{pages}{2265--2270} (\bibinfo{year}{2005}).

\bibitem{Nowak2014}
\bibinfo{author}{I.~Nowak} et al., \enquote{\bibinfo{title}{{Analytical and numerical approach in the simple modelling of thermoacoustic engines}}}, \bibinfo{journal}{Int. J. Heat Mass Tran.} {\textbf{\bibinfo{volume}{77}}}, \bibinfo{pages}{369--376} (\bibinfo{year}{2014}).

\bibitem{ANSYS2012_NavierStokes}
\bibinfo{author}{ANSYS}, \enquote{\bibinfo{title}{{Governing equations}}}, in
  \emph{\bibinfo{booktitle}{ANSYS CFX-Solver Theory Guide}},
  \bibinfo{pages}{18--26} (\bibinfo{publisher}{SAS IP, Inc.})
  (\bibinfo{year}{2012}).

\bibitem{VanderPoel2013}
\bibinfo{author}{B.~{Van der Poel}}, \enquote{\bibinfo{title}{{Time-domain
  impedance boundary conditions in computational fluid dynamics for use in
  thermoacoustic modeling}}}, \bibinfo{type}{MSc. thesis},
  \bibinfo{school}{University of Twente} (\bibinfo{year}{2013}).

\bibitem{Huber2008}
\bibinfo{author}{A.~Huber}, \bibinfo{author}{P.~Romann}, and
  \bibinfo{author}{W.~Polifke}, \enquote{\bibinfo{title}{{Filter-based
  time-domain impedance boundary conditions for CFD applications}}}, in
  \emph{\bibinfo{booktitle}{ASME Turbo Expo 2008: Power for Land, Sea and Air}}
  (\bibinfo{publisher}{ASME}, \bibinfo{address}{Berlin, Germany}), \bibinfo{pages}{1--11}, 
  (\bibinfo{year}{2008}).

\bibitem{Kaess2008}
\bibinfo{author}{R.~Kaess}, \bibinfo{author}{A.~Huber}, and
  \bibinfo{author}{W.~Polifke}, \enquote{\bibinfo{title}{{A time-domain
  impedance boundary condition for compressible turbulent flow}}}, in
  \emph{\bibinfo{booktitle}{14th AIAA/CEAS Aeroacoustics Conference (29th AIAA
  Aeroacoustics Conference)}}, \bibinfo{pages}{1--15}
  (\bibinfo{publisher}{AIAA}, \bibinfo{address}{Vancouver, British Columbia,
  Canada}), \bibinfo{pages}{1--15}, (\bibinfo{year}{2008}).
  
\bibitem{Tijdeman1975}
\bibinfo{author}{H. Tijdeman},
  \enquote{\bibinfo{title}{{On the propagation of sound waves in cylindrical tubes}}}, \bibinfo{journal}{J. Sound Vib.} {\textbf{\bibinfo{volume}{39}}}, \bibinfo{pages}{1--33} (\bibinfo{year}{1975}).
  
\bibitem{Ohmi1982}
\bibinfo{author}{M. Ohmi} and \bibinfo{author}{M. Iguchi},
  \enquote{\bibinfo{title}{{Critical Reynolds number in an oscillating pipe flow}}}, \bibinfo{journal}{B. JSME} {\textbf{\bibinfo{volume}{25}}}, \bibinfo{pages}{165--172} (\bibinfo{year}{1982}).
  
\bibitem{Antao2013}
\bibinfo{author}{D.~S. Antao} and \bibinfo{author}{B. Farouk}, \enquote{\bibinfo{title}{{High amplitude nonlinear acoustic wave driven flow fields in cylindrical and conical resonators}}}, \bibinfo{journal}{J. Acoust. Soc. Am.} {\textbf{\bibinfo{volume}{134}}}, \bibinfo{pages}{917--932} (\bibinfo{year}{2013}).

\bibitem{Swift2002_acousticpower}
\bibinfo{author}{G.~W. Swift}, \enquote{\bibinfo{title}{{Acoustic power}}}, in
  \emph{\bibinfo{booktitle}{Thermoacoustics: A Unifying Perspective for Some
  Engines and Refrigerators}}, \bibinfo{pages}{105--116}
  (\bibinfo{publisher}{Acoustical Society of America}) (\bibinfo{year}{2002}).
\end{thebibliography}
\end{document}